# How can we combat online misinformation? A systematic overview of current interventions and their efficacy

Pica Johansson*[1], Florence Enock*[1], Scott Hale[1,2,3], Bertie Vidgen[4**], Cassidy Bereskin[2**], Helen Margetts[1,2], Jonathan Bright[1]

[1] Public Policy Programme, The Alan Turing Institute, The British Library, 96 Euston Road, London. NW1 2DB.

[2] Oxford Internet Institute, University of Oxford, 1 St Giles', Oxford. OX1 3JS.

[3] Meedan, San Francisco, California, USA

[4] Rewire (https://rewire.online/)

*Equal contribution

** Formerly Online Safety Team, Public Policy Programme, The Alan Turing Institute

Corresponding authors:

Pica Johansson: pjohansson@turing.ac.uk

Florence Enock: fenock@turing.ac.uk

# Abstract

The spread of misinformation is a pressing global problem that has elicited a range of responses from researchers, policymakers, civil society and industry. Over the past decade, these stakeholders have developed many interventions to tackle misinformation that vary across factors such as which effects of misinformation they hope to target, at what stage in the misinformation lifecycle they are aimed at, and who they are implemented by. These interventions also differ in how effective they are at reducing susceptibility to (and curbing the spread of) misinformation. In recent years, a vast amount of scholarly work on misinformation has become available, which extends across multiple disciplines and methodologies. It has become increasingly difficult to comprehensively map all of the available interventions, assess their efficacy, and understand the challenges, opportunities and tradeoffs associated with using them. Few papers have systematically assessed and compared the various interventions, which has led to a lack of understanding in civic and policymaking discourses. With this in mind, we develop a new hierarchical framework for understanding interventions against misinformation online. The framework comprises three key elements: Interventions that *Prepare* people to be less susceptible; Interventions that *Curb* the spread and effects of misinformation; and Interventions that *Respond* to misinformation. We outline how different interventions are thought to work, categorise them, and summarise the available evidence on their efficacy; offering researchers, policymakers and practitioners working to combat online misinformation both an analytical framework that they can use to understand and evaluate different interventions (and which could be extended to address new interventions that we do not describe here) and a summary of the range of interventions that have been proposed to date.

# Introduction

The prevalence of misinformation online is a pressing societal problem that has gained significant attention amongst researchers, policymakers and the wider public. It has been described as a crisis demanding urgent action (Farkas & Schou, 2019), largely because of the multiple efforts to use misinformation to manipulate public opinion (Lewandowsky et al., 2017), the fast pace at which it can spread (Vosoughi et al., 2018) and the lack of transparency in platform governance. Advances in technology, such as the development of large language models and the generation of 'deepfakes', mean that the creation of misinformation is becoming more complex, and tackling it is increasingly difficult to keep pace with. Importantly, the spread of online misinformation has had significant negative consequences in the offline world too. Recently, misinformation has been pointed to as a major force influencing individuals' voting behaviours (Meredith & Morse, 2015); encouraging people to put themselves at severe risk of physical harm (e.g. drinking bleach to supposedly combat illness) (Dharawat et al., 2022); and even risking the lives of others through vigilante mobs (Banaji & Bhat, 2019) or baseless conspiracies (e.g. 'Pizzagate') (Bleakley, 2021).

Because of the many actors implicated by the proliferation of misinformation online and increasing public pressure to intervene, a large network has emerged where different actors may intervene in different ways. There are now critical questions about which actors in society should be most responsible for tackling misinformation (i.e., government, educators and civil society), along with how much of the responsibility should fall on these actors. In this work, we present a unified framework for understanding and classifying these diverse efforts. While misinformation may be propagated offline as well as online, the present work focuses on combating misinformation that is primarily created and spread online via websites and social media platforms.

Despite being a widely used term, misinformation is difficult to define. The term may be understood differently by different people and across different disciplines (Camargo & Simon, 2022; Vraga & Bode, 2020). The most common terms used to describe false information are 'fake news', 'misinformation' and 'disinformation'. Although definitions of disinformation and misinformation were provided in early scholarship on the Internet (e.g., Hernon, 1995), research related to misinformation around the 2016 U.S. presidential election momentarily used the term 'fake news' to describe false articles that were mimicking the presentation of traditional news outlets (Allcott & Gentzkow, 2017; Lazer et al., 2018; Pennycook et al., 2018). Since then, many researchers have shed the term 'fake news', and use misinformation or disinformation, given the imprecision and politicisation of the 'fake news' (e.g., Ecker et al., 2022).

Hernon (1995) defined disinformation as a 'deliberate attempt to deceive or mislead' and misinformation as 'an honest mistake' (p. 134). Wardle and Derakhshan (2017) further formalised and popularised the distinction in their information disorder typology. Under this framework, both mis- and disinformation involve the sharing of false information, but disinformation presumes an 'intent to deceive' behind the propagation or purveying of false information and misinformation, on the other hand, refers to false or misleadingly framed information that is shared unwittingly without intentionality.

Misinformation can refer to many forms of content, including text, image, audio and multimodal. Examples include fake headlines and news articles, fully synthetic photos or videos ('deepfakes'), partially fabricated content such as edited images or content presented outside of their true context known as 'cheap fakes' (Paris & Donovan, 2019). In this work, we use the term misinformation to cover all categories of online false information and do not presuppose an intent to deceive. Furthermore, while misinformation may originate from multiple actors, the interventions that we focus on in this review are primarily designed to tackle online misinformation originating from non-state actors.

Since the first coordinated efforts to spread misinformation online, many interventions have been developed across multiple disciplines, including psychology, cognitive science, education and computer science. Platforms have also implemented new interventions targeting misinformation, and there is growing public pressure to legislate against misinformation. Policies in Spain, Brazil and France target election misinformation and while few countries have reached policies which outright forbid misinformation, legislative measures are stepping up significantly in East and Southeast Asia (Funke & Flamini, 2022; Kajimoto, 2018), such as the 2019 Protection from Online Falsehoods and Manipulation Act (POFMA) in Singapore, that can result in imprisonment and/or large fines for knowingly spreading false information (*Singapore Fake News Laws*, 2022).

There are many interventions to tackle misinformation, spanning educational, cognitive and platform-level approaches.[1] However, many of these interventions are so far limited in their adoption and it is unclear which interventions are most effective. Additionally, it is difficult to compare across the various interventions as it is argued that misinformation research lacks an integrated approach (Guay et al., 2022). Recently, in addition to noting the knowledge gaps on efficacy, researchers have also pointed to the difficulty in assessing the impact and influence of interventions, the practical challenges many interventions face, as well as their ethical implications (Roozenbeek, Suiter, et al., 2022a).

Given the vast (and growing) diversity of potential interventions against misinformation, and the increasing pressure to act, we aim to provide a high-level overview of current approaches to tackling misinformation online. We identify each type of available intervention against misinformation and we position these into a novel collectively exhaustive analytical framework. We also consider the stage at which the interventions are applied, and who is best placed to implement each one. We analyse the interventions comparatively, focusing on discussions around their efficacy and practical feasibility.

# Methods

Our methodological approach is as follows. First, we conducted a comprehensive literature search primarily using relevant search terms, as well as our knowledge of work by established academics in the field in order to identify as many high quality papers as possible relating to misinformation interventions. We considered each type of intervention across several factors, including *how* the intervention is thought to operate (whether educative, cognitively-focused, altering exposure or increasing understanding of key topics); *when* the intervention is applied (before, during or after exposure), *who* is best placed to implement the intervention (e.g., educators, governments, platforms or other organisations); and *which effects* of misinformation the intervention aims to ameliorate (e.g., susceptibility, sharing, other behaviours).

---

[1] For a useful summary, see p.76 of The Department for Digital, Culture, Media and Sport's (DCMS) media literacy strategy:
https://assets.publishing.service.gov.uk/government/uploads/system/uploads/attachment_data/file/1004233/DCMS_Media_Literacy_Report_Roll_Out_Accessible_PDF.pdf

We used these analyses to develop a framework outline in which we propose three main stages of intervention: Prepare, Curb and Respond, which correspond approximately to the timepoints of before, during and after misinformation exposure. Within these three key stages, we propose different levels of operation. These different levels convey both how the interventions are thought to work and who is best placed to implement them. Each intervention type was placed under a different level of the framework. The framework was designed such that it was possible to include all types of misinformation intervention that we identified, and also such that each intervention type could be positioned in only one part of the framework (i.e., collectively exhaustive and mutually exclusive).

Following the creation of our framework outline, we then conducted further literature searches to find examples of how each intervention type has been tested and implemented (if at all) in order to assess the overall efficacy of each one. We used this part of the research process to also critically assess the practical feasibility of the implementation of the interventions and to consider the ethics of each approach. We summarise our framework and provide short descriptions and evaluations of each intervention type in the results section below.

It is worth noting that in this work, we do not intend to provide an exhaustive overview of the misinformation literature. Rather, we choose representative research studies for each intervention and make sure that we reflect recent advances in the field. In doing so, we hope to provide a clear, high-level overview of what has been offered to tackle misinformation, and which approaches are currently most promising.

# Results

## 1.   Defining the stages of intervention

Based on our systematic review of interventions aimed at tackling misinformation online, we identified three key stages at which intervention may be directed.

Firstly, several interventions are aimed at *preparing* people to deal with misinformation before exposure may occur (the 'Prepare' stage of the framework). Interventions at the Prepare stage aim to cognitively equip people with the skills to discern misinformation from factual content, along with providing reminders about the prevalence of misinformation. Within the Prepare stage there are two levels: Educate and Prime. Interventions at the Educate level aim to teach people strategies to recognize misinformation well before exposure and include media literacy courses and inoculation games. Interventions at the Prime level remind people to be alert to misinformation and include public awareness campaigns and general warnings. Interventions at the Prepare stage are typically implemented by nonprofit and civil society organisations, by governments, and in educational settings.

Secondly, many interventions are aimed at *curbing* the spread of misinformation closer to or during the stage of exposure (the Curb stage of the framework). Interventions at the Curb stage aim to limit people's exposure to misinformation by reducing its creation, spread and perceived credibility. Within the Curb stage there are three levels: Contextualise, Slow and Remove. Interventions at the Contextualise level aim to provide additional information alongside certain pieces of content so that users are aware of the wider context before believing or sharing the content. These intervention types include fact-check labels, prompts about checking content before sharing, and provenance cues. Interventions at the Slow level aim to systematically reduce the visibility and spread of misinformation in order to limit people's exposure. These intervention types include making content classed as misinformation unsearchable, demonetising the creation and sharing of such content, and using algorithms to downrank or exclude such content from recommendations. Interventions at the Remove level take down content classed as misinformation or accounts creating such content in order to prevent exposure to the content and to prevent repeated generation of similar content. These intervention types include 'deplatforming' particular users, channels or forums; preventing certain content from being uploaded; and removing certain pieces of content after upload. Interventions at the Curb stage are typically implemented by online services and platforms.

Thirdly, some interventions are aimed at *responding* to misinformation after exposure (the 'Respond' stage of the framework). Interventions at the Respond stage primarily aim to correct false beliefs brought about by exposure to misinformation. Interventions at the Respond stage include Post hoc Corrections, which can take the form of Debunking and Counterspeech, aiming to counter misinformation claims with facts. Often, the aim is to prevent harmful behaviours arising as a result of false beliefs induced by misinformation, and to prevent further spread of such beliefs online and offline. Interventions at the Respond stage are typically implemented by nonprofit and civil society organisations, governments, influencers and individual users.

Table 1, below, shows where each intervention type is placed within the framework and provides a brief definition of each. More detailed descriptions of each intervention, along with examples of their implementation and high-level summaries of their efficacy, are presented in our analysis section below.

**Table 1: A framework for understanding the stages at which current misinformation interventions are targeted.**

| | Intervention Stage, Level and Type | Definition |
|---|---|---|
| 1. | **Prepare** | At the Prepare stage, interventions aim to reduce susceptibility to misinformation by cognitively preparing people for possible exposure. |
| 1.1 | Educate users | At the Educate level, interventions aim to teach people strategies to recognize misinformation well before exposure. |
| 1.1.1 | Media literacy courses | Media literacy courses aim to equip people with the skills necessary to critically evaluate content, recognise content that may be misinformation, and reduce susceptibility to believing and sharing such content. |
| 1.1.2 | Inoculation games | 'Inoculation' against misinformation is designed to reduce people's susceptibility through 'prebunking' - showing people common signs of misinformation through short games. |
| 1.2 | Prime users | At the Prime level, interventions aim to remind people to be alert to recognizing misinformation nearer to the point of exposure. |
| 1.2.1 | General warnings | Generalised misinformation warnings are messages reminding people of the dangers of misinformation and encouraging users to actively scrutinise content. |
| 1.2.2 | Public awareness campaigns | Public awareness campaigns aim to raise public awareness about the prevalence of misinformation and the harm that believing and sharing such content can cause. |
| 2. | **Curb** | At the Curb stage, platforms aim to limit people's exposure to misinformation by reducing its creation and spread. |
| 2.1 | Contextualise content | At the Contextualise level, platforms provide people with additional context about particular pieces of content, aiming to help them make informed assessments about content veracity. |
| 2.1.1 | Fact-check labels | Fact-check labels partially or fully overlay content and warn users that claims made in the content have been disputed, sometimes offering users links to more information. |
| 2.1.2 | Tiplines and self-help resources | Self-help resources aim to allow individuals to investigate the veracity of a claim or gather additional context, for example through social media tiplines, bots and fact-check databases. |
| 2.1.3 | Prompts | Prompts are used to encourage people to pause before liking or sharing content to consider its veracity, for example by asking people if they would like to read a full article before sharing a headline. |
| 2.1.4 | Provenance cues | Provenance cues provide information about the source and edit history of audio-visual online content (its metadata), to help people understand if something is presented out of context or is a deepfake. |
| 2.2 | Slow content | At the Slow level, platforms make content classed as misinformation less visible and de-incentivise the creation and sharing of such content. The aim is to limit people's exposure to misinformation. |
| 2.2.1 | Demonetisation | Demonetising content means that publishers of misinformation cannot make money from it, for example by generating an ad-revenue. |
| 2.2.2 | Algorithmic downranking | Algorithmically downranking content limits a piece of content's amplification on the platform or service. The content may appear less frequently, be shown to fewer users, or will appear further down a 'feed' or list of recommendations. |
| 2.2.3 | Delisting | De-listing content means that it does not appear in results when using terms or hashtags, but the content remains accessible on the service.  In some instances, this may be implemented on the search query level such that chosen queries return no results. |
| 2.3 | Remove users and/or content | At the Remove level, platforms remove content or creators of content that is classed as misinformation. The aim is to prevent exposure to misinformation entirely. |
| 2.3.1 | Early stage moderation | Early stage moderation involves blocking content at the point of upload to prevent certain content from ever appearing on the platform. The aim is to prevent exposure completely. |
| 2.3.2 | Deplatforming | Deplatforming removes a user, channel or forum from a platform when they post content classed as misinformation (or violating other rules of the platform). The aim is to prevent generation of further misinformation from the same source. |
| 3. | **Respond** | At the Respond stage, interventions aim to correct false beliefs induced as a result of exposure to misinformation. |
| 3.1 | Correct claims post hoc | Post hoc corrections are interventions providing corrections about claims made in content classed as misinformation, often aiming to prevent harmful behaviours arising as a result of exposure, and limiting further spread. |
| 3.1.1 | Debunking | Debunking interventions aim to correct false beliefs brought about as a result of misinformation exposure by countering claims made in misinformation with detailed factual explanations. |
| 3.1.2 | Counterspeech | Counterspeech comprises user-generated rebuttals which aim to challenge claims made in misinformation, for example in comments under particular pieces of content. |

# 2. Assessing the efficacy and feasibility of each intervention type

In the following section, we provide a high level summary of each of the interventions presented in the framework and draw on scholarly literature to examine the efficacy of each one. We discuss the practical and ethical feasibility of each intervention type and, where possible, we provide examples of the implementation of each. The intervention types are discussed section-by-section in the same order as presented in Table 1.

## 1. Prepare

### 1.1 Educate

#### 1.1.1 Media literacy courses

Often, educational interventions against misinformation form broader digital or media literacy programmes which aim to provide people with analytical skills to help them navigate the large volumes of information online (Pappas, 2022). These courses broadly aim to equip people with knowledge, tools and strategies for effectively assessing the veracity of online content. For example, some parts of these courses may show people examples of misinformation in order to demonstrate common tactics used by those who create and share misleading and fabricated content.

One prominent example of an educational intervention is Learn to Discern (*Learn to Discern (L2D) - Media Literacy Training | IREX*, n.d.), launched in 2018 by the International Research and Exchanges Board (IREX). Learn to Discern is a media literacy training curriculum for educators and students. The training programme includes units on, for example, recognising types of misinformation such as clickbait, along with how to verify content sources. In Ukraine between 2015 and 2016, the programme equipped over 400 grassroots trainers to teach 15,000 citizens and also trained over 1,100 teachers in Ukraine who in turn taught the curriculum to 7,500 students. Another similar example is the News Literacy Project (www.newslit.org) a US-based nonpartisan, not-for-profit organisation, which aims to provide free programmes and resources for both educators and individuals to teach, learn and share news literacy skills. Resources include e-learning platform 'Checkology', podcasts, newsletters and blogs.

The efficacy of courses such as the ones described above is promising. Impact studies have shown that the Learn to Discern programme is effective with a diverse range of participants. In the case study in Ukraine, 18 months after receiving training, adults were still 25% more likely to report checking multiple news sources, and 13% more likely to correctly identify a fake news story. School-aged children were 18% better at identifying fake news stories (Murrock et al., 2018). Similar success of the programme has been reported through a case study in Jordan (*Case Study: Learn to Discern in Jordan*, n.d.). However, outcomes may differ depending on context. Some research shows that education programmes are particularly effective for assessing breaking news events and evaluating partisan content (Bulger & Davison, 2018).

A recent large-scale study also found promising effects of a real-world digital media literacy intervention in both the United States and India (Guess et al., 2020). The intervention significantly improved discernment between mainstream and false news headlines among a nationally representative sample in the US and among a highly educated sample in India. However, there were no effects among a representative sample of respondents in a largely rural area of northern India, where rates of social media use are lower. Further, the increased discernment only remained measurable several weeks later in the sample from the United States, not in the sample from India. Generally, however, academic studies testing media literacy initiatives outside the west remain scarce (Roozenbeek, Suiter, et al., 2022b).

Media literacy courses are broadly feasible given the body of free resources online, and are primarily implemented by schoolteachers and specialist trainers. Across the globe, several countries have implemented national media literacy campaigns, including Finland, Australia, Belgium, Canada, Denmark, Nigeria, Singapore and the Netherlands (Funke & Flamini, 2022). As these interventions work by teaching people how to navigate their online environments, they do not raise ethical questions surrounding interference with the online environment as might be the case with particular forms of moderation that remove or suppress specific content. However, media literacy campaigns backed by authoritarian regimes raise other ethical questions. In addition, some researchers note that courses such as these are not without limitation: they require voluntary uptake and may struggle to reach populations with low digital media literacy, such as older individuals (Bronstein & Vinogradov, 2021).

Furthermore, some research suggests that media literacy endows people with a false sense of confidence in their ability to critically assess media (Bulger & Davison, 2018). Research should continue to measure the long-term effectiveness of educational courses, with a particular focus on informing the design of such interventions for optimal reach and efficacy in the future.

### 1.1.2 Inoculation games

Inoculation games are designed to reduce people's susceptibility to misinformation through 'prebunking'. In these games, players typically navigate a simulated social media environment while also learning about common methods that purveyors of misinformation rely on. Players may also be exposed to common cues which signal content as misinformation, such as emotionally manipulative language, incoherence, false dichotomies, scapegoating, and ad hominem attacks. The process of psychological 'inoculation' is rooted in the idea that just as injections containing a weakened strain of a virus trigger antibodies in the immune system to help confer resistance against future infection, the same can be achieved in building protection against misinformation. By preemptively warning people against misleading tactics and by exposing people to a weakened version, it is claimed that cognitive resistance can be developed against a range of misinformation types and contexts (Lewandowsky & van der Linden, 2021; Traberg et al., 2022). While the games currently rely on voluntary participation, it is possible that they could be implemented in educational settings, or be routinely offered by platforms and websites as users are browsing.

Several inoculation games against misinformation have been developed, including Bad News (www.getbadnews.com), Go Viral! (www.goviralgame.com) and Harmony Square (https://harmonysquare.game/en). Bad News was developed by researchers at Cambridge university in collaboration with Dutch media company 'DROG' (Basol et al., 2020). In the game, players take on the role of a 'fake news tycoon', aiming to gather as many followers as possible while maintaining credibility. The game is designed to expose common manipulation techniques that are used to mislead people online. For example, players can earn badges corresponding to tactics used in the production of fake news, such as taking advantage of political polarization, employing conspiracy theories, and discrediting other sources. Go Viral! was developed by the same researchers at Cambridge in partnership with the UK Cabinet Office to specifically tackle misinformation about COVID-19 . The game demonstrates how misinformation can sow doubt through the use of fake experts and emotionally charged language.

Many studies support the efficacy of inoculation games for reducing susceptibility to misinformation (Roozenbeek & van der Linden, 2019), boosting people's confidence in their ability to detect misinformation (Basol et al., 2020), and reducing misinformation sharing intentions (Roozenbeek & Linden, 2020). These 'inoculation effects' have been replicated in a non-Western sample (Iyengar et al., 2022), and there is also evidence to show that effects last for up to two months with 'booster' doses (Maertens et al., 2021).

Owing to the apparent success of inoculation games, recent work has examined how psychological inoculation against misinformation may be delivered efficiently through other mediums. One recent paper found that in a field study conducted on YouTube (n = 22,632), watching a short inoculation animation about manipulation techniques commonly used in misinformation improved manipulation technique recognition, boosted people's confidence in spotting these techniques, increased people's ability to discern trustworthy from untrustworthy content, and improved the quality of their sharing decisions (Roozenbeek, van der Linden, et al., 2022).

Inoculation shows promise: a number of empirical studies support their efficacy and these findings are replicated across different samples. The games are feasible to implement given that they are typically free to access online, and they are designed to be fun and relatively fast to complete. As the games foster critical thinking skills prior to misinformation exposure, these interventions avoid ethical concerns surrounding content moderation. However, the games require voluntary adoption and it is currently unclear how high uptake is. Research also shows that these games require specificity to be effective, which remains a challenge as it is hard to predict which kinds of misinformation people will be exposed to (Zerback et al., 2021). Additionally, some researchers suggest that the games may not enhance discrimination between true and false content as much as is claimed, and rather elicit more conservative responding overall such that people are more sceptical of true content as well (Batailler et al., 2022; Modirrousta-Galian & Higham, 2022).

Further research should focus on discrimination between true and false content as an outcome measure to ensure that inoculation games do not simply enhance scepticism overall. Further work should also aim to understand the long-term effectiveness of inoculation games and, if promising, develop ways to encourage uptake within schools, places of work and as part of media literacy initiatives.

## 1.2 Prime

### 1.2.1 General warnings

General misinformation warnings are typically messages found in the online environment reminding people about the prevalence of misinformation and advising people to stay alert to possible exposure (Clayton et al., 2020; Greene & Murphy, 2021). For example, Google sometimes places warnings above unreliable search results that typically arise in breaking news situations and also in more general situations where it determines search results are likely to be low quality (Robertson, 2021). The warnings are presented at the top of the search page and aim to manage expectations about the reliability of the search results. General warnings about misinformation could also be implemented on social media platforms and other websites.

Overall support for the efficacy of generalised warnings for combating misinformation is weak. Greene & Murphy (2021) failed to find any effect for  general warnings on the negative outcomes of misinformation exposure. Here, participants read short vignettes relating to COVID-19, four true and four false. For example, a false vignette might describe how people who drink more than three cups of coffee a day experience COVID-19 symptoms more mildly. Participants were then asked about related behavioural intentions, for example, whether they planned to drink more coffee. Beforehand, participants received either a positive misinformation warning (warning framed in terms of gain), a negative misinformation warning (warning framed in terms of loss), no misinformation warning, or no public health message. Results showed that viewing a general warning about misinformation, whether positively or negatively framed, had no impact on participants' acceptance of misinformation from fake news, nor on behavioural intentions, and there was no effect of warning condition on truthfulness ratings for any of the stories.

Research offering marginally more support for the efficacy of generalised warnings comes from Clayton and colleagues (2020). Results from a large online experiment indicated that false headlines are perceived as somewhat less accurate when people receive a general warning about misleading information on social media and this effect did not vary by whether headlines were congenial to respondents' political views. However, exposure to a general warning also reduced belief in the accuracy of true headlines as well as false ones, suggesting that efforts to promote scepticism toward false news may have spillover effects and also decrease trust in legitimate news and information: See Pantazi and colleagues (2021) for an in-depth review of the tension between excess gullibility and excess vigilance in the context of misinformation.

While generalised warnings may be a feasible and low-cost solution to implement, on the whole, findings suggest general warnings are ineffective (Greene & Murphy, 2021) or are considerably less effective than other interventions (Clayton et al., 2020; Ecker et al., 2010). As a result, resources for combatting online misinformation should be directed at other approaches.

### 1.2.2 Public awareness campaigns

Public awareness campaigns are broadly aimed at increasing awareness of the prevalence of online misinformation across society. These campaigns are typically implemented by governments and platforms and take short, advert-like formats.

The UK government's 'Take care with what you share' campaign (www.sharechecklist.gov.uk) includes a short video which outlines key indicators of misinformation and encourages people to pause before liking, commenting or sharing online content. The associated SHARE checklist urges people to consider the Source of the information; to read beyond the Headline only (reading articles in full and checking dates); to Analyse the content by using fact-checking services; to look out for whether images and videos have been Retouched or edited (using reverse image search if needed); and whether there are any obvious Errors in the content such as bad grammar, which could indicate an untrustworthy source. Similarly, the Be Media Smart campaign (www.bemediasmart.ie) in Ireland was developed by members of Media Literacy Ireland (www.medialiteracyireland.ie) to encourage people to Stop, Think and Check when consuming and interacting with online content.

More globally, 'Stop the Spread' (*Countering Misinformation about COVID-19*, 2019) was launched in May 2020 as a collaboration between the World Health Organization (WHO) and the UK government with the aim to raise awareness about the risks of misinformation around COVID-19. Promoted across Africa, Asia, Europe, Middle East and Latin America, the campaign encouraged people to check and verify information about COVID-19 with trusted sources such as the WHO and other health authorities. The campaign ran on BBC World News and BBC.com.

The extent to which public awareness campaigns reduce susceptibility to (and sharing of) misinformation is unclear. As these campaigns are intended to be generalised and targeted across society, research has not measured their efficacy in a controlled setting. However, other pre-bunking interventions are shown to have high efficacy (see sections 1.1.1 and 1.1.2 on media literacy courses and inoculation games, above), which may suggest public awareness campaigns have the potential to be beneficial. However, such campaigns do not teach critical tools for navigating the online environment in as much depth as education interventions. In this sense, awareness campaigns may be more akin to a scaled version of generalised warnings, for which efficacy is weak (see section 1.2.1, above).

Public awareness campaigns are broadly feasible to implement and have the potential for a wide reach given they can take the form of paper leaflets and posters (e.g., distributed in libraries), television and radio ads, and social media campaigns. As the campaigns are designed to increase alertness to possible misinformation before exposure, they avoid ethical concerns surrounding content moderation. However, one concern that public awareness campaigns have in common with generalised warnings is that in promoting scepticism, audiences may also experience decreased trust in legitimate news stories (Clayton et al., 2020; Pantazi et al., 2021). Additionally, the efficacy of public awareness campaigns relies on trust in whichever authority implements the campaign, which may differ heavily between individuals (J. D. West & Bergstrom, 2021).

Further research should focus on discrimination between true and false content as an outcome measure to ensure that the kinds of messages found in public awareness campaigns do not simply enhance scepticism overall. Such campaigns may be best used in conjunction with deeper educational approaches, serving as important reminders for people to draw on the media literacy skills they previously learned.

## 2. Curb

### 2.1 Contextualize

#### 2.1.1 Fact-check labels

Fact-check labels are full or partial overlays which typically warn users that claims made in particular pieces of content have been disputed by fact-checkers, often offering links to more information about the topic. Fact-check labels are usually based on the judgements of expert human fact-checkers. These labels are most commonly implemented by social media platforms and search engines.

There is overall support for the efficacy of fact-check labels in reducing susceptibility to misinformation and reducing misinformation sharing intentions (for a review, see Nieminen & Rapeli, 2019). One research study showed that adding a 'disputed' or 'rated false' tag to a headline significantly lowered its perceived accuracy relative to a control condition and that these overlays were more effective at reducing susceptibility to misinformation than general warnings (Clayton et al., 2020). Additionally, in the same study, fact-check labels did not affect the perceived accuracy of unlabeled false or true headlines (unlike general warnings mentioned above). Other studies similarly find that exposure to a fact-check tag improves accuracy judgments about the specific content (Ecker et al., 2010; Nyhan et al., 2020).

There is some evidence to suggest that the time point at which fact-check labels are shown may be important for their effectiveness. In one study, participants read headlines taken from social media and saw 'true' or 'false' tags either before, during or after exposure. Accuracy judgements about the headlines one week later were highest when participants had seen the fact-checks after headline exposure, as opposed to before or during (Brashier et al., 2021).

One issue often noted with fact-check labels is that they rely on the judgements of professional fact-checkers. The process is laborious and human checkers struggle to keep up with the enormous amount of content posted on social media each day. Additionally, individual fact-checkers may be (or may be perceived as being) biased or politically motivated in their assessments. Addressing this issue, recent work suggests that fact-checking may be reliably crowdsourced without impairing the quality of judgments. One recent paper showed that average veracity ratings of politically balanced groups of laypeople correlate highly with judgments of professional fact-checkers, suggesting that employing a 'wisdom of the crowds' approach is a promising way to enhance scalability and reduce perceived bias in fact-checking interventions (Allen et al., 2021). Twitter's 'Birdwatch' (Coleman, 2021) implements this crowd-sourced approach. Here, members of the programme write notes contextualising posts or provide related information about certain pieces of content. According to Twitter, once a Birdwatch note is attached to a tweet, users are 15 to 35 percent less likely to engage with it compared to users who aren't shown the note (Kelly, 2022). Birdwatch, however, is on a volunteer basis and contrasts with Allen et al. (2021), who used politically-balanced crowds. In contrast, annotations on Birdwatch are highly polarised (Yasseri & Menczer, 2022).

Overall, there is evidence to suggest that fact-check labels can successfully enhance accuracy judgements about online content. They are relatively straightforward to implement and are likely to be perceived as a less severe intervention than full content removal. However, one unintended consequence of fact-checking is the 'implied truth effect', whereby false headlines that fail to get tagged are considered validated and are seen as more accurate (Pennycook, Bear, et al., 2020). Additionally, the efficacy of fact-check labels relies on users' trust of the decision-makers.

Recent work shows promise in using Natural Language Processing (NLP) to match social media content with already fact-checked content (Kazemi et al., 2021, 2022; Shaar et al., 2020). This method allows for scalable cross-referencing, as a more efficient way to fact-check than doing so post-by-post. In Kazemi et al. (2022) the classification and retrieval experiments were conducted in monolingual, multilingual, and cross-lingual settings, achieving 86% average accuracy for match classification.

As fact-checking is a time-consuming and under-resourced process, future work will benefit from further understanding the overall efficacy of crowd-generated fact-checks and NLP-based solutions for a more scalable, credible, and less costly solution.

### 2.1.2 Tiplines and self-help resources

It is not always possible to tag content with fact-check labels. Discovering misinformation content to label and labelling that content is challenging on end-to-end encrypted platforms such as WhatsApp, LINE, VIber, and Signal, as the platform operator does not have visibility into the content. It is technically feasible to store hashes or fingerprints of known misinformation content locally on a smartphone or tablet and attach fact-checking labels when content matches the hashes (e.g., Kazemi et al., 2022; Reis et al., 2020). However, despite no data leaving the device, such approaches are generally unpopular at present with the Electronic Frontier Foundation (EFF) among others opposing them as being a slippery slope, an invasion of privacy, or a weakening of encryption (Portnoy, 2019). Such systems could potentially uphold user privacy and provide warnings about misinformation similar to how antivirus software works on most computers, but it would be essential that the user be fully in control of what scanning happens, what actions are taken, and that no data leave the device without the users' explicit permission. More human-centred design research is needed in this area.

For now, WhatsApp and other platforms have started misinformation 'tiplines.' These are accounts on social media platforms to which a user can forward a message that possibly contains misinformation and discover fact-checks and other misinformation resources from independent fact-checking organisations (*IFCN Fact-Checking Organizations on WhatsApp | WhatsApp Help Center*, n.d.). Analysing one such tipline in India that operated during the country's 2019 general elections, Kazemi et al. (2022) found the tipline captured the most popular content being shared in large, "public" WhatsApp groups and that such content was captured quickly: messages were often shared with the WhatsApp tipline before appearing in the large WhatsApp groups the authors were data mining.

In September and October 2022, five fact-checking organisations and technology non-profit Meedan partnered with Brazil's election authority (the TSE) to run a WhatsApp bot that similarly allowed users to ask questions, forward potential misinformation posts, and discover fact-checks. The bot was heavily promoted by WhatsApp and used by millions of Brazilian citizens (*Meedan Launches Collaborative Effort to Address Misinformation on WhatsApp during Brazil's Presidential Election*, 2022). Other self-help resources include Google's Fact-Check Explorer that aggregates publicly available fact-checks and makes them searchable (toolbox.google.com/factcheck/explorer). Computer science researchers have also built prototype systems to analyse claims, search existing fact-checks, search the web for content that supports or refutes claims, or perform other tasks that could assist a user in analysing the veracity of a claim (Chernyavskiy et al., 2021; Hassan et al., 2017; Panayotov et al., 2022; Smeros et al., 2021).

Additional resources include the labels WhatsApp applies to messages forwarded many times and the affordance it offers to run a Google search using the highly-forwarded message as a query (*About Forwarding Limits | WhatsApp Help Center*, n.d.). LINE, an end-to-end encrypted messaging app popular in Taiwan and Japan, also implements a misinformation tipline approach, using one central, 'official' account to operate the tipline rather than the accounts of individual fact-checkers as WhatsApp does (Deck & Elliott, 2021). LINE also shares debunks of prevalent misinformation claims on its Line Today homepage within the app. Academic research on tiplines is nascent but likely of growing importance as more communication moves to end-to-end encrypted messaging apps.

### 2.1.3 Prompts

Prompts that appear when a user is about to post content (also referred to as 'accuracy nudges' or 'friction') are designed to draw a user's attention to the accuracy of a headline prior to them sharing it, or otherwise slow down human interaction with a platform (Kirchner & Roesner, 2022). These prompts are typically found on social media platforms. The shift in focus intends to induce extra caution or to make users think twice prior to sharing. The intervention targets the act of sharing misinformation, shown to substantially reduce its reach (Andı & Akesson, 2021; Pennycook et al., 2021). The appeal of posting prompts are that they are proactive, as well as that they allow full freedom for users to decide for themselves what content to publish and remove technology companies from the position of having to decide what is true or false.

Pennycook and Rand's (2022) recent internal meta-analysis using 20 experiments (N = 26,863) showed that asking users to consider the accuracy of content prior to sharing reduced the intention to share false headlines by 10% relative to the control, thereby increasing the quality of news people shared (Pennycook & Rand, 2022). These results were not moderated by factors such as race, gender or education, and held across various topics - indicating that the findings are generalizable and replicable. This intervention gets at what the authors have evidenced previously, namely that the problem is not that individuals struggle to tell true headlines from false ones, but that there could be additional distraction at the point of sharing. Therefore, drawing attention to accuracy changes sharing intentions (Pennycook, McPhetres, et al., 2020). Fazio (2020 came to a similar conclusion in their study, which asked people to pause for a few seconds to consider the accuracy of a post before sharing it.

Research on posting prompts has inspired multiple campaigns such as the United Nations misinformation initiative, "Pause" ('Pause before Sharing, to Help Stop Viral Spread of COVID-19 Misinformation', 2020), which encourages users to pause before sharing content relating to Covid-19. Platforms have also been seen to have been using this intervention in various contexts. For example, Twitter prompts users to read articles if they haven't opened the link prior to retweeting. Twitter later reported that this initiative had resulted in users opening articles 40% more often when having received this message (Hutchinson, 2020). The following year, Facebook followed suit and added a similar prompt (Spangler, 2021). WhatsApp has added friction by limiting the number of times a message can be forwarded as well as the number of people that can be in one group.[2]

Although studies indicate that nudges can be a nonintrusive and simple way to curtail misinformation shared on social media, these results have been contested (Roozenbeek et al., 2021). In addition, there is a lack of studies aiming to understand how widespread this intervention could be, and whether the effects will be watered down with repeated use. Researchers elsewhere have also raised concern that  low-cost nudges might displace support for high-cost measures, such as policy interventions (Kozyreva et al., 2020). The intervention does benefit from being specific in its target and point of intervention, but the longer term efficacy is questionable and remains understudied.

---

[2] At the time of writing, WhatsApp limits groups to a maximum of 512 people and allows messages to be forwarded to a maximum of five groups at once. If a message has been forwarded at least five times, it can only be forwarded to one additional group at a time. Further details are at:
https://blog.whatsapp.com/reactions-2gb-file-sharing-512-groups and
https://faq.whatsapp.com/1053543185312573/?locale=en_US

2.1.4 Provenance cues

Provenance information documents the origins of audio and visual online content. Provenance technology for tackling misinformation works by using advances in cryptographic digital signatures to track the origins of audiovisual content (the metadata), including the time, date, authorship and location of creation, as well as any manipulation that has occurred during the course of its distribution (discussed in Gregory, 2022). The technology is designed for use by primary media sources to enhance trust and verification in mainstream journalism.

Provenance cues detailing factors such as image source and history would typically be presented alongside audio or visual content online. Because the focus is on verifying the source of audiovisual content, provenance technology is particularly aimed at forms of misinformation which rely on recontextualising content. Examples of this include content which presents out of context images to support fake headlines, such as viral images purporting to show crisis actors in Ukraine applying fake blood to their faces as claiming that the war in Ukraine is a hoax (Sardarizadeh & Robinson, 2022) (in fact the images were from a 2020 TV set). Other examples include the use of 'cheapfakes', manipulating audiovisual content with software such as photoshop to support false claims, and 'deepfakes', creating purely synthetic content made to look real, again to support false claims (Dan et al., 2021).

The idea of using provenance cues to tackle 'out of context' online misinformation is, at the time of writing, relatively new, and there are currently no large-scale research studies that test the efficacy of provenance cues in reducing susceptibility to misinformation. However, development of the intervention is being supported by a number of projects such as the Coalition for Provenance and Authenticity (C2PA) (www.c2pa.org), the Content Authenticity Initiative (CAI) (www.contentauthenticity.org), and Project Origin (www.originproject.info) (see also The News Provenance Project - www.newsprovenanceproject.com).

Because provenance technology relies on cryptography to provide users with metadata about the source of online content, this form of intervention does not rely on humans in the way that, for example, fact-check labels do. In this way, provenance cues have the potential to be more scalable than fact-check labels, and could also be perceived as more credible. However, the efficacy of the intervention will rely on user engagement—people will need to critically assess whether the claims presented in a piece of content match the corresponding provenance information. Additionally, trust in provenance cues may differ between users. While little is currently known about how social media users might respond to provenance technologies or how these might be implemented in practice, this is a potentially rich area for future research.

## 2.2 Slow

### 2.2.1 Demonetisation

Scholars have found that internet intermediaries such as Google and Facebook act as enablers for fake news publishers, as those who intend to profit from their content are motivated by extending the reach of their content (Nizamani, 2020). Demonetising misinformation is an intervention aimed at weakening the financial incentives associated with the dissemination of misinformation, and is therefore an intervention best actioned by platforms and search engines Misinformation often appears as 'clickbait' and are known for their virality, their existence is often financially motivated as those creating the misinformation can game the platforms sensationalist preference for financial gain (Hughes & Waismel-Manor, 2021; Verstraete et al., 2021). The problem also extends to misinformation served as advertisement and hoax websites. In their study sampling 20,000 illegitimate new sites, the nonprofit Global Disinformation Index found that ad technology companies spend about $235 million annually by running ads on these types of websites (*The Quarter Billion Dollar Question: How Is Disinformation Gaming Ad Tech?*, 2019).

Facing immense scrutiny after the misinformation campaigns published in the lead-up to the 2016 US presidential election, platforms have taken action to minimize the financial incentives to spread misinformation. Google have updated their ad policy to reflect a ban on misinformation concerning Covid-19 (Dang, 2020), as well as climate misinformation (Hiar, 2021). Ahead of elections held in the United States (2020), Germany (2021) and France (2022), Meta announced a number of policies focusing specifically on securing the integrity of elections, such as banning ads that delegitimise details of the vote or undermine voter safety (*Information on Prohibited Ads Related to Voting and Ads about Social Issues, Elections or Politics.*, n.d.). Despite these initiatives, curbing the spread of misinformation remains a challenge across a range of languages, as recently highlighted in the run-up to Brazil's 2022 General Election ('Facebook Fails to Tackle Election Disinformation Ads Ahead of Tense Brazilian Election', 2019).

Beyond the platforms themselves, there are also calls for ad-networks to do more to reduce the monetary incentives to spread misinformation (D. M. West, 2017). One study found that of the top-10 credible ad-servers, those that liaise between retailers and websites selling ad-space account for 66.7% of fake ad traffic (Bozarth & Budak, 2021). The researchers also noted how these ads account for a small proportion of their revenue and in a separate study find that the purchases are placed on a small number of websites. As Ellie Vorhaben points out in the Chicago Policy Review, these actors are in a position where cutting these ties would be both low effort, and have minimal financial impact. Vorhaben also notes that there is also space for more work from the ad-purchasing side, with large corporations still placing ads on illegitimate news sites, thereby making the case that both retailers and consumers could do more (Vorhaben, 2022).

Regardless of who acts on it, demonetisation presents itself as an appealing solution for a range of actors, as it sidesteps the arguments often concerned about freedom of speech, instead targeting the incentives that have allowed the misinformation industry to flourish. However, one of the key difficulties of its implementation revolves around effectively identifying information at scale, especially in an industry where many adverts are bought and sold automatically.

2.2.2 Algorithmical downranking

In efforts to avoid interfering with freedom of expression there have been calls for increased focus on adapting platforms' algorithms, the mechanism for ranking the 'feed' of content, or recommendations, to which a user is exposed. Certain algorithms control how content is presented to users, downranking content therefore means to modify how often and to how many the content appears (Saltz & Leibowicz, 2021). This intervention makes a distinction between the right to have certain content published and its amplification; the right to publish content remains, but there is no 'right to reach'. This is especially important in the case of misinformation which, it has been suggested, is often more successful at achieving high position in many social media ranking algorithms (Shin & Valente, 2020), or is likely to be recommended to users even if they have not shown a prior interest in such content (O'Callaghan et al., 2015).

Algorithmic downranking is currently one of the most commonly used interventions by platforms. For example, YouTube downranks unauthoritative content (Courchesne et al., 2021) and Facebook downranks exaggerated or sensationalist health claims (Perez, 2019) Content may even be ranked to zero, meaning it has no ranking and will therefore not be algorithmically delivered to other users in the feed, but it will remain on the platform (Saltz & Leibowicz, 2021). In 2018, Facebook claimed that its downranking efforts cut future views by more than 80% on average for posts that had been labelled as 'false' by third-party fact-checkers (Lyons, 2018).

Despite being a strategy commonly deployed by platforms, algorithmic downranking remains understudied. Data access for researchers is limited meaning that a true understanding of the workings of algorithms is usually difficult to establish. Unfortunately, the proprietary algorithms underpinning platforms is also an important intellectual property, meaning there are strong limitations on whom it can be shared with. Lacking this information, scholars have instead undertaken studies that contribute to how the signals that feed into ranking algorithms might be improved. Work in this area tests to what degree crowdsourcing can be utilised, such as if users could help make judgement calls on sources (Epstein et al., 2020) or on specific pieces of content (Roitero et al., 2021).

Epstein and colleagues (2020) asked users (N = 984) to rate the trustworthiness of sources, participants believing that their results would influence ranking algorithms. Based on previous work, the authors speculated that activating users' reasoning would decrease their susceptibility to misinformation (Bago et al., 2020; Martel et al., 2020). Results showed that the crowd ratings could efficiently identify misinformation sources, and that the minimal partisanship polarisation of results seen were equal (therefore cancelling one other out). Further research is needed, but these results show that platforms may want to consider the role of users in helping supply information for downranking algorithms.

### 2.2.3 Delisting

Another option available to platforms to keep unwanted content at bay is de-listing it. Removing content from any search results provided by the platform means that users who are not specifically looking for misinformation or already involved in communities that spread conspiracies are less likely to find it. Such actions can also include removing hashtags which are another common mechanism of content discovery. For example, Pinterest blocked search results for anti-vaccine terms even before the COVID-19 pandemic (Telford, 2019). There have been multiple instances of platforms banning hashtags that are associated with specific misinformation campaigns, such as those related to the conspiracy that the 2020 US election was "stolen" (Perez & Hatmaker, 2020) and hashtags related to coronavirus misinformation (Jin, 2020). Similar to other interventions discussed in this work, de-listing content is an option that benefits from upholding freedom of expression whilst limiting users access to misinformation and therefore limiting its reach.

There are currently no studies that discuss the efficacy of de-listing content. However, there are doubts among scholars who study other forms of online harms, such as communities that promote eating disorders or extremist views. This research points to the fact that users evade policing of the platforms by using alternative terms and hashtags, or co-opt seemingly mundane hashtags and instead use coded language (Chancellor et al., 2016; Gerrard, 2018; Lakomy, 2022). Based on these studies, it is reasonable to assume that when misinformation is suppressed by making it unsearchable, these narratives too find ways to stay on platforms by making use of alternative language that has yet to be detected. Furthermore, removing hashtags may also stifle legitimate expression on some topics. For example, a report by UNESCO showed that, in addition to banning many conspiracy associated hashtags, Tiktok had also removed the hashtag 'holocaust'; which may interrupt misinformation but also interrupts legitimate discourse (*History under Attack: Holocaust Denial and Distortion on Social Media*, 2022) .

## 2.3 Remove

### 2.3.1 Early stage moderation

Early stage moderation involves blocking or removing content from a social media platform either as it is being uploaded or shortly afterwards. Current moderation solutions for misinformation rely heavily on reporting mechanisms and human-led moderation. However, post-by-post moderation based on individual users is not a scalable solution, and research has found that reporting mechanisms more generally are fairly under-used. Despite this, platforms have made certain efforts to increase user-reporting of misinformation. For example, Twitter has tested a "report misleading content" button ('Twitter Tests "misleading"

Post Report Button for First Time', 2021), and introduced its crowdsourced 'Birdwatch' programme in early 2021, in which users taking part can assess the accuracy of posts.

As early stage moderation interventions sit with the platforms, the evidence base to evaluate post-by-post moderation remains thin. However, several researchers have attempted to explore how platforms could intervene at the point of upload. For example, some research examines how analysing the contents of a post could help to stop the spread of misinformation. In a study conducted by Zhou and colleagues (2020), news content was mined for attributes at four language levels: lexicon-level, syntax-level, semantic-level and discourse-level. At each level the researchers outlined misinformation 'indicators', such as if the articles have click-bait related attributes, what terms they use, sentence length and part-of-speech tagging. When all attributes are taken into account they could predict misinformation with 88% accuracy. Castelo and colleagues (2019) produce similar results in their work, which also includes web mark-up features such as the number of advertisements on a page.

In many instances, large misinformation campaigns have been traced back to activity by algorithmically driven social media accounts, also known as bots. Research shows that the curbing of bots could be an effective strategy for mitigating the spread of misinformation in early stages, as a large proportion of the total traffic that carries misinformation can be traced back to relatively few accounts (Shao et al., 2018; Vosoughi et al., 2018). In their analysis of 14 million messages, Shao and colleagues (2018) find that roughly one-third of "low-credibility" content is spread by only 6% of accounts. However, there is also an emphasis on the fact that bots alone do not explain the success of misinformation, as humans in many cases amplify the spread of misinformation at the same rate as bots do (Vosoughi et al., 2018).

Any automated solution risks miscategorising accounts, although the identification of bot accounts in order to slow the spread of misinformation is a relatively low-risk strategy to adapt. It may also be the case that bot accounts or networks become increasingly sophisticated and therefore manage to go undetected—making this strategy difficult to maintain long term.

Predicting the probability that content is misinformation before it starts to propagate on social media, either by scanning content, links or propagation networks may rely on an AI based solution. While such solutions do scale effectively to the volume of content on social media, they also create further risks, such as a potential difficulty for users in terms of understanding why content has been deleted and the potential for incorrectly deleting accurate content. Hence, caution is needed alongside robust monitoring and appeals mechanisms for any such automated systems.

A final removal avenue that will likely be explored over the coming years is how and whether there are policy avenues that could ban misinformation, as scholarly work has made the case for characterising misinformation as online advertising fraud (Braun & Eklund, 2019). Current legislation underway in the UK that could have provisions in place for this kind of activity include the Online Advertising Programme (*Online Advertising Programme*, 2022) consultation, as well as the Online Safety Bill (*Online Safety Bill - Parliamentary Bills - UK Parliament*, 2022) – the latter which recently included fraud as a priority offence.

### 2.3.2 Deplatforming

Deplatforming refers to the removal, ban or suspension of a user from a social media platform (Rogers, 2020). It can be seen as a de-facto boycotting of the person in question by removing their ability to contribute to the platform. While platforms routinely delete large volumes of user accounts, the intervention is also often associated with high-profile cases and as an intervention used only as a last resort, and in many cases as a reaction to public pressure, as platforms firmly stand-by their position of not interfering with free speech. Moreover, it is not in their interest to remove these personalities with large followings who create viral content, due to the financial implications of doing so. In terms of platform adoption, another aspect to take into account is how platforms go about de-platforming. Facebook was for example criticised in 2019 when it announced the removal of two far-right influencers prior to their ban, allowing them to usher followers to other platforms (Martineau, 2019).

Aside from deplatforming individuals, platforms can also do sweeps of de-platforming users who are spreading disinformation. Twitter, for example, removed 70,000 QAnon accounts after the storming of the US Capitol (Conger, 2021). Finally, deplatforming can also encompass the shutting down of whole forums, such as the subreddit r/fatpeoplehate (Chandrasekharan et al., 2017).

De-platforming is seen as a severe measure which many researchers have taken interest in (albeit not only from the perspective of misinformation). Research mainly concern tracking the progression of topics and reach of the de-platformed interest (Rauchfleisch & Kaiser, 2021), activity of their supporters (Jhaver et al., 2021), any potential backlash or counter-reactions (Innes & Innes, 2021) and migration of follower bases (Bryanov et al., 2022; Ribeiro et al., 2021; Rogers, 2020).

A study by Jhaver and colleagues (2021) found that deplatforming significantly disrupted discussions about 'influencers' that are deplatformed. Their analysis, based on 49m Tweets collected six months before and after the deplatforming of Alex Jones, Milo Yiannopoulos and Owen Benjamin, also concluded that de-platforming significantly reduced the popularity of many of the anti-social ideas associated with the influencers. A small group of supporters did however increase their activity in reaction to deplatforming—consistent with other findings which show that removal might have negative counter reactions both on the platform in question, and across the wider ecosystem (Ali et al., 2021).

Other research on de-platforming shows the opposite effect. A study by Innes and Innes (2021) collected data mentioning two Covid-19 conspiracy influencers, QAnon affiliated David Icke and Kate Shemirani. Icke was de-plaformed from Facebook in April 2020 because of repeatedly spreading Covid-19 misinformation. Their research showed that the removal attracted additional attention to the influencer, and in the 7 days following his account removal his mentions on Facebook increased by 84%. Informed by empirical analysis, their study proceeds to conceptualise two possible reactions to de-platforming, the creation of so-called "minion accounts" and general efforts to "re-platform". 'Minion accounts' are accounts that surface after deplatforming and which have a clear association with the removed 'leader'. The accounts continue to post in promotion of the mission or message or the leader, although not under any direction. The emergence of 'minion accounts' is one of many 're-platforming' responses, showing persistent diversification of strategies to disseminate their ideas such as diversifying their cross-platform presence.

Despite much talk of migration of alt-platforms, few studies quantify these movements. An analysis by Rauchfleisch and Kaiser (2021) found that of the 516 far-right YouTube channels analysed in their 2018–2019 study, 111 had been de-platformed of which 20 could be found on BitChute. They concluded that deplatforming was effective in minimising the reach of misinformation in YouTube, and that despite some users flocking to alternative platforms they cannot make-up for the loss in viewership, which is consistent with other quantitative findings on that their audiences on new platforms ultimately 'thin' (Rogers, 2020). Despite having smaller audiences, others point to how deplatforming can have the tendency to harden the views of followers and those engaging with the misinformation (Dwoskin & Timberg, 2021) or how the act of deplatforming can, at least temporarily, bring more attention to them—the 'Streisand effect' (Innes & Innes, 2021).

Existing studies on deplatforming are difficult to compare for many reasons, but the effects on various platforms are also found to differ (for research on Twitter, see Jhaver et al. (2021); Facebook, see Innes & Innes (2021); YouTube, see Klinenberg (2022) and Rauchfleisch & Kaiser (2021); and Reddit, see Chandrasekharan et al. (2017)). Rogers' (2020) study of how de-platformed accounts move from various social media platforms to Telegram concludes that the intervention helps 'clean-up' platforms such as Facebook and Instagram, but is less effective on YouTube and Twitter. There also remains differences in the nature of de-platforming, such as the systematic removal of networked accounts spreading misinformation vs individual "influencer" accounts. Findings also differ depending on how much time, post-deplatformisation, the research draws conclusions based on. Most research, however, does indicate that de-platforming users in one way or another curbs momentum, limits content virality and decreases the reach of de-platformed users/groups. The intervention seems to have an impact in terms of short-term disruption but places platforms in the position of de-facto arbiters of free speech, without eradicating the problem or necessarily increasing the 'health' of social media.

# 3. Respond

## 3.1 Correct post hoc

### 3.1.1 Debunk

Debunking interventions aim to correct false beliefs brought about as a result of misinformation exposure by countering claims made in misinformation with detailed factual explanations. The process of debunking often aims to prevent further spread of false beliefs both online and offline, also aiming to prevent harmful actions (and inactions) that may arise as a result of such false beliefs, such as taking unsafe treatments for illness or refusing vaccines (for a summary, see Roozenbeek & van der Linden, 2022).

It is important to note the distinction between debunking interventions and fact-check labels. While both depend on the process of fact-checking, fact-check labels overlay content at the point of exposure and prompt the user to engage with caution. Debunking approaches, however, provide facts about claims made in prevalent pieces of misinformation, explaining why claims may be misleading or untrue, and replacing falsehoods with accurate information about the topic.

Debunking efforts may be implemented by governments, the media, nonprofits, individual influencers and other organisations, and may be disseminated through radio and newspaper reports, television documentaries, and social media. For example, Telltale Research (www.telltaleresearch.com) recently collaborated with influencer Abigail Thorn to create a video debunking myths surrounding the dangers of the Covid-19 vaccine for her channel Philosophy Tube, gathering over one million views in 8 months. YouTubers such as cardiologist Rohin Frances, with 500,000 subscribers, create content to educate the public about medicine and debunk common healthcare related misinformation.

There is evidence to suggest that post-hoc debunking works to reduce belief in misinformation. Work on the 'continued inference effect', whereby beliefs in falsehoods persist even after they are corrected, showed that continued false beliefs after misinformation exposure are less likely to persist if a retraction is accompanied by an alternative account as opposed to presenting just a retraction. This shows that providing facts and explanations about false or misleading claims is likely to be more effective at lowering beliefs than just adding 'disputed' tags or retracting pieces of content (Ecker et al., 2010; Paynter et al., 2019). A meta-analysis also found that the debunking effect was weaker when the debunking message simply labelled misinformation as incorrect rather than when it introduced corrective information (Chan et al., 2017). However, another study found that a real word debunking campaign only reduced belief in misinformation in participants with low or medium level beliefs, not those with high beliefs in the misinformation content. Further, debunking did not affect behavioural intentions (Helfers & Ebersbach, 2022), nor do changes in beliefs brought about by debunking necessarily last (Paynter et al., 2019).

Given the nuances identified surrounding the efficacy of debunking for misinformation, researchers have recently put together a series of debunking handbooks which outline best practices for post hoc corrections about misinformation (*The Debunking Handbook*, 2020). For example, one section explains how to optimise effective debunking for health practitioners, suggesting leading with the factual content, mentioning a myth only once and with a warning, explaining exactly why and how the myth misleads, and then finishing with factual information.

Note that while there has previously been concern about possible 'backfire effects' resulting from debunking efforts—the idea that repeating falsehoods may strengthen belief in them even in the context of debunking (Ecker et al., 2020)—more recent research finds no reliable evidence for these, showing the risk of side effects from debunking is low (Swire-Thompson et al., 2020).

Taken together, evidence for the efficacy of post-hoc debunking correcting false beliefs induced by misinformation exposure is promising. Such interventions have the potential for a wide reach because they can be implemented through various outlets. Additionally, such interventions attempt to facilitate education and understanding surrounding complex topics, meaning effects may be more durable than simply taking content down. However, some researchers suggest that debunking efforts may not reach as many people as the initial misinformation (Kostygina et al., 2020) and the efficacy of post hoc debunking depends on how trusted the source is (Ecker & Antonio, 2021) as well as trust in institutions and media more broadly (Jamieson et al., 2021). While debunking alone is unlikely to be sufficient for tackling the spread of misinformation, it is an important part of a multi-pronged approach. To enhance benefits, efforts could be spent targeting factual explainers where they are most needed (for example, specific social media groups and forums), and making sure post hoc corrections take the optimal format.

### 3.1.2 Counterspeech

Counterspeech is a way to counter potentially harmful content online by presenting an alternative viewpoint. The key idea behind counterspeech is to encourage a greater amount of (positive) speech online rather than attempting to mitigate harmful content through removal (Benesch, 2014; Buerger, 2021a)[3]. Up until now, research into counterspeech has tended to focus on tackling hate speech (as opposed to misinformation). However, theory suggests that the principles may be applied to misinformation as well as hate speech, and indeed the two harms often overlap (Cinelli et al., 2021; Hameleers et al., 2022).

While it is currently difficult to estimate how many users engage in counterspeech online, or how many real world instances of counterspeech are generated across platforms, there have been some high profile movements which employ this approach. The Sweden-based #jagärhär ('#Iamhere') Facebook group has over 70,000 members who take part in coordinated responses to hateful comments online. This model has spread across several other countries, together now forming 'I am Here International' (https://iamhereinternational.com/) - a global citizen-based network tackling online hostility through counterspeech (discussed in Buerger, 2021). Relatedly, the Online Civil Courage Initiative (OCCI) (*Online Civil Courage Initiative (OCCI)*, n.d.) aims to teach users how to use counterspeech to push back against harmful content onlines.

While no studies to our knowledge have directly tested the efficacy of counterspeech for ameliorating the negative effects of online misinformation exposure, recent experimental research suggests that some forms of counterspeech may be effective for tackling hate speech (e.g., Hangartner et al., 2021; Munger, 2017). In work by Hangartner and colleagues (2021), English-speaking Twitter users (N=1350) who had previously sent hateful content were randomly assigned to one of three counterspeech strategies—empathy-based, warning-based, humour-based, or a control. Results showed that being exposed to empathy-based counterspeech increased the retrospective deletion of hate speech and reduced the future creation of hate speech over a 4-week follow-up period relative to the control. The other kinds of counterspeech did not yield the same positive results. The results show that the specific content included in counterspeech is likely to play a key role in its efficacy. However, it is important to note that this study measured effects on hate speech and not misinformation and may not be generalisable. Indeed, work elsewhere in the misinformation literature suggests that simple rebuttals are less effective than offering detailed alternative narratives for countering misinformation (Clayton et al., 2020; Ecker et al., 2010; Ecker et al., 2022).

---

[3] For multiple resources, see: https://dangerousspeech.org/counterspeech/

Given the magnitude of harmful misinformation online, counterspeech may be beneficial in helping to tackle the problem at scale. While this type of intervention does not rely on platform uptake, little is known about how many users create and engage with counterspeech, or how achievable it would be to encourage more to do so. Bot-generated counterspeech offers one solution to the practical feasibility of this intervention (e.g., de los Riscos & D'Haro, 2021). First, though, more empirical research is needed to assess how well counterspeech works in reducing the spread of false beliefs brought about by misinformation, and which type of counterspeech would be most effective in this context.

# Discussion and conclusion

This work has provided a systematic overview of the interventions offered to tackle misinformation online, all presented within a framework which identifies three key stages at which an intervention may be directed. Interventions at the Prepare stage show promise, in particular at the Educate level, with digital media literacy interventions showing effective outcomes, as well as inoculation games and videos that may be faster to implement. The support for interventions at the Prime level is weaker, but these could be folded into deeper educational approaches, for example as upkeep or boosts for digital media literacy initiatives.

Among the interventions at the Curb stage aimed at stopping the spread of misinformation, many show promise for becoming more widespread in use. For example, at the Contextualise level, provenance cues are a particular area worth dedicating new research efforts into examining their efficacy and feasibility. The most extensively researched intervention at this level, fact-check labels, might be effective for reducing sharing intentions, but their long-term efficacy is contested. At the Slow level, the wider adoption of policy efforts to demonetise misinformation will target important system-level change. At the Block level, deplatforming is effective in minimising reach and virality, but the act of removing users remains controversial and might have unintended consequences. Across the interventions aiming to Curb or limit the spread of misinformation, platform buy-in remains a key challenge.

Of the interventions which at the Respond stage, post hoc corrections, particularly debunking, continue to play a key role in the life cycle of combating misinformation. In particular, these efforts are particularly effective if they replace a false narrative with an alternative one, and if this correction comes from a trusted source. While debunking alone is unlikely to be sufficient for tackling the spread of misinformation, and preventing initial susceptibility to misinformation through education is likely to be more efficient than correcting false beliefs once they are in place, post hoc corrections continue to be an important last defence within a multi-pronged approach.

In its approach to join up empirical findings with considerations around the feasibility and ethics of various interventions, this review has also identified a discrepancy between the interventions currently being used by platforms, and those studied by social science researchers. We echo Courchesne and colleagues' (2021) call to shift the research focus to algorithmic downranking, content removal and deplatforming, as well as making it a priority to focus more on real world testing. This is no easy task, but there are positive indicators that research opportunities are opening up, for example under provisions that are currently underway in the Digital Services Act, as well as other landmark legislation. In line with these discussions, we too encourage formalised pathways for platforms and researchers to collaborate on real-world testing.

This review attempts to give a broad overview of the interventions proposed to tackle misinformation online, but its breadth also comes with limitations. The review is not exhaustive and remains high-level, although specific attention has been made to include key pieces of scholarship that are most representative of the field, and therefore we hope the references included will help readers to seek out further information. Follow up work would benefit from including an analysis of the many policy options currently being debated to combat misinformation, as well as systemic factors such as the decreasing trust in media.

We also acknowledge that it is difficult to compare across types of misinformation, for example the topic (e.g. health misinformation vs. political misinformation) and content type (e.g. text vs. video) (for a discussion on this point, see Guay et al., 2022). More generally, the field's ability to identify fruitful strategies to tackle misinformation is severely undermined by the lack of an integrated approach for evaluating interventions across studies. We therefore welcome the framework proposed by Guay and colleagues (2022), which unifies the discussion based on common research designs, and encourages researchers to use this framework to develop a field-wide best practice for evaluating misinformation interventions. As research on the efficacy of interventions progresses, special attention should be paid to assessing any potential adverse consequences. For example, it is important to make sure that interventions only increase scepticism to misinformation, not all information (Basol et al., 2020; Modirrousta-Galian & Higham, 2022; Pantazi et al., 2021; Roozenbeek, van der Linden, et al., 2022; Roozenbeek & van der Linden, 2019).

Finally, more research is needed in non-Western contexts. As articulated by Roozenbeek and colleagues (2022) the geographic concentration of research on misinformation severely limits what we know about the phenomenon and the intervention strategies. Early research comparing interventions across countries shows that initiatives that work in one context may not be receiving the same results elsewhere, leaving substantial gaps in our claims of understanding the efficacy of even the most promising interventions (e.g., Guess et al., 2020). Meanwhile, research also continues to investigate whether individual differences could be a better way to compare efficacy than across demographics (Arechar et al., 2022).

Combating misinformation is likely to require a combination of different approaches in order to fully address the problem. Those that spread misinformation or seek to benefit from it are likely to adapt, despite interventions, meaning approaches to combat misinformation will also need to evolve and remain active. The prospect of an increased pace and sophistication of the dissemination of misinformation is of great concern; we therefore reiterate the importance of involving the entire information ecosystem. In providing a new analytical framework for understanding the full range of misinformation interventions, we open a discussion on the feasibility, ethics and efficacy of each type, laying important groundwork for those taking practical steps to combat the creation, proliferation and consequences of harmful misinformation.

# References

*About forwarding limits | WhatsApp Help Center*. (n.d.). Retrieved 21 December

2022, from https://faq.whatsapp.com/1053543185312573/?locale=en_US

Ali, S., Saeed, M. H., Aldreabi, E., Blackburn, J., De Cristofaro, E., Zannettou, S.,

& Stringhini, G. (2021). Understanding the Effect of Deplatforming on

Social Networks. *13th ACM Web Science Conference 2021*, 187–195.

https://doi.org/10.1145/3447535.3462637

Allcott, H., & Gentzkow, M. (2017). Social Media and Fake News in the 2016

Election. *Journal of Economic Perspectives*, *31*(2), 211–236.

https://doi.org/10.1257/jep.31.2.211

Allen, J., Arechar, A. A., Pennycook, G., & Rand, D. G. (2021). Scaling up

fact-checking using the wisdom of crowds. *Science Advances*, *7*(36),

eabf4393.

Andı, S., & Akesson, J. (2021). Nudging Away False News: Evidence from a

Social Norms Experiment. *Digital Journalism*, *9*(1), 106–125.

https://doi.org/10.1080/21670811.2020.1847674

Arechar, A. A., Allen, J. N. L., Berinsky, A., Cole, R., Epstein, Z., Garimella, K.,

Gully, A., Lu, J. G., Ross, R. M., Stagnaro, M., Zhang, J., Pennycook, G., &

Rand, D. (2022). *Understanding and Combatting COVID-19 Misinformation*

*Across 16 Countries on Six Continents*. PsyArXiv.

https://doi.org/10.31234/osf.io/a9frz

Bago, B., Rand, D. G., & Pennycook, G. (2020). Fake news, fast and slow:

Deliberation reduces belief in false (but not true) news headlines. *Journal*

*of Experimental Psychology. General*, *149*(8), 1608–1613.

https://doi.org/10.1037/xge0000729

Banaji, S., & Bhat, R. (2019, November 11). WhatsApp Vigilantes: An exploration

of citizen reception and circulation of WhatsApp misinformation linked to mob violence in India. *Media@LSE*.

https://blogs.lse.ac.uk/medialse/2019/11/11/whatsapp-vigilantes-an-explora
tion-of-citizen-reception-and-circulation-of-whatsapp-misinformation-linked-
to-mob-violence-in-india/

Basol, M., Roozenbeek, J., & van der Linden, S. (2020). Good News about Bad
News: Gamified Inoculation Boosts Confidence and Cognitive Immunity
Against Fake News. *Journal of Cognition*, *3*(1), 2.
https://doi.org/10.5334/joc.91

Batailler, C., Brannon, S. M., Teas, P. E., & Gawronski, B. (2022). A Signal
Detection Approach to Understanding the Identification of Fake News.
*Perspectives on Psychological Science: A Journal of the Association for
Psychological Science*, *17*(1), 78–98.
https://doi.org/10.1177/1745691620986135

Benesch, S. (2014). Countering Dangerous Speech: New Ideas for Genocide
Prevention. *SSRN Electronic Journal*. https://doi.org/10.2139/ssrn.3686876

Bleakley, P. (2021). Panic, pizza and mainstreaming the alt-right: A social media
analysis of Pizzagate and the rise of the QAnon conspiracy. *Current
Sociology*, 00113921211034896.
https://doi.org/10.1177/00113921211034896

Bozarth, L., & Budak, C. (2021). Market Forces: Quantifying the Role of Top
Credible Ad Servers in the Fake News Ecosystem. *Proceedings of the
International AAAI Conference on Web and Social Media*, *15*, 83–94.
https://doi.org/10.1609/icwsm.v15i1.18043

Brashier, N. M., Pennycook, G., Berinsky, A. J., & Rand, D. G. (2021). Timing
matters when correcting fake news. *Proceedings of the National Academy
of Sciences*, *118*(5), e2020043118.
https://doi.org/10.1073/pnas.2020043118

Braun, J. A., & Eklund, J. L. (2019). Fake News, Real Money: Ad Tech Platforms,

Profit-Driven Hoaxes, and the Business of Journalism. *Digital Journalism*, *7*(1), 1–21. https://doi.org/10.1080/21670811.2018.1556314

Bronstein, M. V., & Vinogradov, S. (2021). Education alone is insufficient to combat online medical misinformation. *EMBO Reports*, *22*(3), e52282.

Bryanov, K., Vasina, D., Pankova, Y., & Pakholkov, V. (2022). The Other Side of Deplatforming: Right-Wing Telegram in the Wake of Trump's Twitter Ouster. In D. A. Alexandrov, A. V. Boukhanovsky, A. V. Chugunov, Y. Kabanov, O. Koltsova, I. Musabirov, & S. Pashakhin (Eds.), *Digital Transformation and Global Society* (pp. 417–428). Springer International Publishing. https://doi.org/10.1007/978-3-030-93715-7_30

Buerger, C. (2021a). *Counterspeech: A Literature Review* (SSRN Scholarly Paper No. 4066882). https://doi.org/10.2139/ssrn.4066882

Buerger, C. (2021b). #iamhere: Collective Counterspeech and the Quest to Improve Online Discourse. *Social Media + Society*, *7*(4), 205630512110638. https://doi.org/10.1177/20563051211063843

Bulger, M., & Davison, P. (2018). The Promises, Challenges, and Futures of Media Literacy. *Journal of Media Literacy Education*, *10*(1), 1–21. https://doi.org/10.23860/JMLE-2018-10-1-1

Camargo, C. Q., & Simon, F. M. (2022). Mis- and disinformation studies are too big to fail: Six suggestions for the field's future. *Harvard Kennedy School Misinformation Review*. https://doi.org/10.37016/mr-2020-106

*Case Study: Learn to Discern in Jordan*. (n.d.). Retrieved 21 December 2022, from https://www.irex.org/project/learn-discern-l2d-media-literacy-training

Castelo, S., Santos, A., Almeida, T., Pham, K., Freire, J., Elghafari, A., & Nakamura, E. (2019). A topic-agnostic approach for identifying fake news pages: 2019 World Wide Web Conference, WWW 2019. *The Web Conference 2019 - Companion of the World Wide Web Conference, WWW 2019*, 975–980. https://doi.org/10.1145/3308560.3316739

Chan, M. S., Jones, C. R., Hall Jamieson, K., & Albarracín, D. (2017). Debunking: A Meta-Analysis of the Psychological Efficacy of Messages Countering Misinformation. *Psychological Science*, *28*(11), 1531–1546. https://doi.org/10.1177/0956797617714579

Chancellor, S., Pater, J. A., Clear, T., Gilbert, E., & De Choudhury, M. (2016). #thyghgapp: Instagram Content Moderation and Lexical Variation in Pro-Eating Disorder Communities. *Proceedings of the 19th ACM Conference on Computer-Supported Cooperative Work & Social Computing*, 1201–1213. https://doi.org/10.1145/2818048.2819963

Chandrasekharan, E., Pavalanathan, U., Srinivasan, A., Glynn, A., Eisenstein, J., & Gilbert, E. (2017). You Can't Stay Here: The Efficacy of Reddit's 2015 Ban Examined Through Hate Speech. *Proceedings of the ACM on Human-Computer Interaction*, *1*(CSCW), 31:1-31:22. https://doi.org/10.1145/3134666

Chernyavskiy, A., Ilvovsky, D., & Nakov, P. (2021). WhatTheWikiFact: Fact-Checking Claims Against Wikipedia. *Proceedings of the 30th ACM International Conference on Information & Knowledge Management*, 4690–4695. https://doi.org/10.1145/3459637.3481987

Cinelli, M., Pelicon, A., Mozetič, I., Quattrociocchi, W., Novak, P. K., & Zollo, F. (2021). Dynamics of online hate and misinformation. *Scientific Reports*, *11*(1), 22083. https://doi.org/10.1038/s41598-021-01487-w

Clayton, K., Blair, S., Busam, J. A., Forstner, S., Glance, J., Green, G., Kawata, A., Kovvuri, A., Martin, J., & Morgan, E. (2020). Real solutions for fake news? Measuring the effectiveness of general warnings and fact-check tags in reducing belief in false stories on social media. *Political Behavior*, *42*(4), 1073–1095.

Coleman, K. (2021, January 25). *Introducing Birdwatch, a community-based approach to misinformation*.

https://blog.twitter.com/en_us/topics/product/2021/introducing-birdwatch-a-

community-based-approach-to-misinformation

Conger, K. (2021, January 12). Twitter, in Widening Crackdown, Removes Over

70,000 QAnon Accounts. *The New York Times*.

https://www.nytimes.com/2021/01/11/technology/twitter-removes-70000-qa

non-accounts.html

*Countering misinformation about COVID-19*. (2019, May 13).

https://www.who.int/news-room/feature-stories/detail/countering-misinform

ation-about-covid-19

Courchesne, L., Ilhardt, J., & Shapiro, J. N. (2021). Review of social science

research on the impact of countermeasures against influence operations.

*Harvard Kennedy School Misinformation Review*.

https://doi.org/10.37016/mr-2020-79

Dan, V., Paris, B., Donovan, J., Hameleers, M., Roozenbeek, J., van der Linden,

S., & von Sikorski, C. (2021). Visual Mis- and Disinformation, Social Media,

and Democracy. *Journalism & Mass Communication Quarterly*, *98*(3),

641–664. https://doi.org/10.1177/10776990211035395

Dang, S. (2020, July 17). Google bans ads on coronavirus conspiracy theory content. *Reuters*.

https://www.reuters.com/article/us-health-coronavirus-advertising-idUSKC

N24I27N

de los Riscos, A. M., & D'Haro, L. F. (2021). ToxicBot: A Conversational Agent to

Fight Online Hate Speech. In L. F. D'Haro, Z. Callejas, & S. Nakamura

(Eds.), *Conversational Dialogue Systems for the Next Decade* (pp. 15–30).

Springer. https://doi.org/10.1007/978-981-15-8395-7_2

Deck, A., & Elliott, V. (2021, March 7). How Line is fighting disinformation without sacrificing privacy. *Rest of World*.

https://restofworld.org/2021/how-line-is-fighting-disinformation-without-sacri

ficing-privacy/

Dharawat, A., Lourentzou, I., Morales, A., & Zhai, C. (2022). Drink Bleach or Do
What Now? COVID-HeRA: A Study of Risk-Informed Health Decision
Making in the Presence of COVID-19 Misinformation. *Proceedings of the
International AAAI Conference on Web and Social Media*, *16*, 1218–1227.
https://doi.org/10.1609/icwsm.v16i1.19372

Dwoskin, E., & Timberg, C. (2021, January 17). Misinformation dropped
dramatically the week after Twitter banned Trump and some allies.
*Washington Post*.
https://www.washingtonpost.com/technology/2021/01/16/misinformation-tru
mp-twitter/

Ecker, U. K. H., & Antonio, L. M. (2021). Can you believe it? An investigation into
the impact of retraction source credibility on the continued influence effect.
*Memory & Cognition*, *49*(4), 631–644.
https://doi.org/10.3758/s13421-020-01129-y

Ecker, U. K. H., Lewandowsky, S., & Chadwick, M. (2020). Can corrections spread
misinformation to new audiences? Testing for the elusive familiarity
backfire effect. *Cognitive Research: Principles and Implications*, *5*(1), 41.
https://doi.org/10.1186/s41235-020-00241-6

Ecker, U. K. H., Lewandowsky, S., Cook, J., Schmid, P., Fazio, L. K., Brashier, N.,
Kendeou, P., Vraga, E. K., & Amazeen, M. A. (2022). The psychological
drivers of misinformation belief and its resistance to correction. *Nature
Reviews Psychology*, *1*(1), Article 1.
https://doi.org/10.1038/s44159-021-00006-y

Ecker, U. K. H., Lewandowsky, S., & Tang, D. T. (2010). Explicit warnings reduce
but do not eliminate the continued influence of misinformation. *Memory &
Cognition*, *38*(8), 1087–1100.

Epstein, Z., Pennycook, G., & Rand, D. (2020). Will the Crowd Game the Algorithm? Using Layperson Judgments to Combat Misinformation on Social Media by Downranking Distrusted Sources. *Proceedings of the 2020 CHI Conference on Human Factors in Computing Systems*, 1–11. https://doi.org/10.1145/3313831.3376232

Facebook fails to tackle election disinformation ads ahead of tense Brazilian election. (2019, August 15). *Global Witness*. https:///en/campaigns/digital-threats/facebook-fails-tackle-election-disinformation-ads-ahead-tense-brazilian-election/

Farkas, J., & Schou, J. (2019). *Post-truth, fake news and democracy: Mapping the politics of falsehood*. Routledge.

Fazio, L. (2020). Pausing to consider why a headline is true or false can help reduce the sharing of false news. *Harvard Kennedy School Misinformation Review*, *1*(2). https://doi.org/10.37016/mr-2020-009

Funke, D., & Flamini, D. (2022). A guide to anti-misinformation actions around the world. *Poynter*. https://www.poynter.org/ifcn/anti-misinformation-actions/

Gerrard, Y. (2018). Beyond the hashtag: Circumventing content moderation on social media. *New Media & Society*, *20*(12), 4492–4511. https://doi.org/10.1177/1461444818776611

Greene, C. M., & Murphy, G. (2021). Quantifying the effects of fake news on behavior: Evidence from a study of COVID-19 misinformation. *Journal of Experimental Psychology: Applied*, *27*(4), 773–784. https://doi.org/10.1037/xap0000371

Gregory, S. (2022). Deepfakes, misinformation and disinformation and authenticity infrastructure responses: Impacts on frontline witnessing, distant witnessing, and civic journalism. *Journalism*, *23*(3), 708–729. https://doi.org/10.1177/14648849211060644

Guay, B., Berinsky, A., Pennycook, G., & Rand, D. (2022). *How To Think About Whether Misinformation Interventions Work*. PsyArXiv.

https://doi.org/10.31234/osf.io/gv8qx

Guess, A. M., Lerner, M., Lyons, B., Montgomery, J. M., Nyhan, B., Reifler, J., & Sircar, N. (2020). A digital media literacy intervention increases discernment between mainstream and false news in the United States and India. *Proceedings of the National Academy of Sciences*, *117*(27), 15536–15545.

Hameleers, M., van der Meer, T., & Vliegenthart, R. (2022). Civilized truths, hateful lies? Incivility and hate speech in false information – evidence from fact-checked statements in the US. *Information, Communication & Society*, *25*(11), 1596–1613. https://doi.org/10.1080/1369118X.2021.1874038

Hangartner, D., Gennaro, G., Alasiri, S., Bahrich, N., Bornhoft, A., Boucher, J., Demirci, B. B., Derksen, L., Hall, A., Jochum, M., Munoz, M. M., Richter, M., Vogel, F., Wittwer, S., Wüthrich, F., Gilardi, F., & Donnay, K. (2021). Empathy-based counterspeech can reduce racist hate speech in a social media field experiment. *Proceedings of the National Academy of Sciences*, *118*(50), e2116310118. https://doi.org/10.1073/pnas.2116310118

Hassan, N., Arslan, F., Li, C., & Tremayne, M. (2017). Toward Automated Fact-Checking: Detecting Check-worthy Factual Claims by ClaimBuster. *Proceedings of the 23rd ACM SIGKDD International Conference on Knowledge Discovery and Data Mining*, 1803–1812. https://doi.org/10.1145/3097983.3098131

Helfers, A., & Ebersbach, M. (2022). The differential effects of a governmental debunking campaign concerning COVID-19 vaccination misinformation. *Journal of Communication in Healthcare*, *0*(0), 1–9. https://doi.org/10.1080/17538068.2022.2047497

Hernon, P. (1995). Disinformation and misinformation through the internet: Findings of an exploratory study. *Government Information Quarterly*, *12*(2), 133–139. https://doi.org/10.1016/0740-624X(95)90052-7

Hiar, C. (2021, October 8). Google Bans Ads That Spread Climate Misinformation. *Scientific American*. https://www.scientificamerican.com/article/google-bans-ads-that-spread-climate-misinformation/

*History under attack: Holocaust denial and distortion on social media*. (2022). UNESCO. https://unesdoc.unesco.org/ark:/48223/pf0000382159

Hughes, H. C., & Waismel-Manor, I. (2021). The Macedonian Fake News Industry and the 2016 US Election. *PS: Political Science & Politics*, *54*(1), 19–23. https://doi.org/10.1017/S1049096520000992

Hutchinson, A. (2020, June 10). Twitter is Adding a New Prompt on Retweets When Users Haven't Opened the Link. *Social Media Today*. https://www.socialmediatoday.com/news/twitters-adding-a-new-prompt-on-retweets-when-users-havent-opened-the-lin/579595/

*IFCN fact-checking organizations on WhatsApp | WhatsApp Help Center*. (n.d.). Retrieved 21 December 2022, from https://faq.whatsapp.com/5059120540855664/?locale=en_US

*Information on prohibited ads related to voting and ads about social issues, elections or politics.* (n.d.). Meta Business Help Centre. Retrieved 21 December 2022, from https://en-gb.facebook.com/business/help/253606115684173

Innes, H., & Innes, M. (2021). De-platforming disinformation: Conspiracy theories and their control. *Information, Communication & Society*, *0*(0), 1–19. https://doi.org/10.1080/1369118X.2021.1994631

Iyengar, A., Gupta, P., & Priya, N. (2022). Inoculation against conspiracy theories: A consumer side approach to India's fake news problem. *Applied Cognitive Psychology*, *n/a*(n/a). https://doi.org/10.1002/acp.3995

Jamieson, K. H., Romer, D., Jamieson, P. E., Winneg, K. M., & Pasek, J. (2021). The role of non–COVID-specific and COVID-specific factors in predicting a shift in willingness to vaccinate: A panel study. *Proceedings of the National Academy of Sciences*, *118*(52), e2112266118. https://doi.org/10.1073/pnas.2112266118

Jhaver, S., Boylston, C., Yang, D., & Bruckman, A. (2021). Evaluating the Effectiveness of Deplatforming as a Moderation Strategy on Twitter. *Proceedings of the ACM on Human-Computer Interaction*, *5*(CSCW2), 381:1-381:30. https://doi.org/10.1145/3479525

Jin, K.-X. (2020, December 18). Keeping People Safe and Informed About the Coronavirus. *Meta*. https://about.fb.com/news/2020/12/coronavirus/

Kajimoto, M. (2018, March 14). In East and Southeast Asia, misinformation is a visible and growing concern. *Poynter*. https://www.poynter.org/fact-checking/2018/in-east-and-southeast-asia-mis information-is-a-visible-and-growing-concern/

Kazemi, A., Garimella, K., Gaffney, D., & Hale, S. (2021). Claim Matching Beyond English to Scale Global Fact-Checking. *Proceedings of the 59th Annual Meeting of the Association for Computational Linguistics and the 11th International Joint Conference on Natural Language Processing (Volume 1: Long Papers)*, 4504–4517. https://doi.org/10.18653/v1/2021.acl-long.347

Kazemi, A., Li, Z., Pérez-Rosas, V., Hale, S. A., & Mihalcea, R. (2022). *Matching Tweets With Applicable Fact-Checks Across Languages* (arXiv:2202.07094). arXiv. https://doi.org/10.48550/arXiv.2202.07094

Kelly, M. (2022, September 7). Twitter is expanding its experimental community moderation system. *The Verge*. https://www.theverge.com/2022/9/7/23341003/twitter-birdwatch-community -moderation-notes-misinformation-tweets-qanon

Kirchner, C., & Roesner, F. (2022). Digital Disinformation: Taxonomy, Impact, Mitigation, and Regulation (Dagstuhl Seminar 21402). *Dagstuhl Reports*, *11*(9), 28–44. https://doi.org/10.4230/DagRep.11.9.28

Klinenberg, D. (2022). *Does Deplatforming Work? Unintended Consequences of banning far-right content creators from social media* (SSRN Scholarly Paper No. 4019767). https://doi.org/10.2139/ssrn.4019767

Kostygina, G., Szczypka, G., Tran, H., Binns, S., Emery, S. L., Vallone, D., & Hair, E. C. (2020). Exposure and reach of the US court-mandated corrective statements advertising campaign on broadcast and social media. *Tobacco Control*, *29*(4), 420–424. https://doi.org/10.1136/tobaccocontrol-2018-054762

Kozyreva, A., Lewandowsky, S., & Hertwig, R. (2020). Citizens versus the internet: Confronting digital challenges with cognitive tools. *Psychological Science in the Public Interest*, *21*(3), 103–156.

Lakomy, M. (2022). Why Do Online Countering Violent Extremism Strategies Not Work? The Case of Digital Jihad. *Terrorism and Political Violence*, *0*(0), 1–38. https://doi.org/10.1080/09546553.2022.2038575

Lazer, D. M. J., Baum, M. A., Benkler, Y., Berinsky, A. J., Greenhill, K. M., Menczer, F., Metzger, M. J., Nyhan, B., Pennycook, G., Rothschild, D., Schudson, M., Sloman, S. A., Sunstein, C. R., Thorson, E. A., Watts, D. J., & Zittrain, J. L. (2018). The science of fake news. *Science*, *359*(6380), 1094–1096. https://doi.org/10.1126/science.aao2998

*Learn to Discern (L2D)—Media Literacy Training | IREX*. (n.d.). Retrieved 21 December 2022, from https://www.irex.org/project/learn-discern-l2d-media-literacy-training

Lewandowsky, S., Ecker, U. K. H., & Cook, J. (2017). Beyond Misinformation: Understanding and Coping with the "Post-Truth" Era. *Journal of Applied Research in Memory and Cognition*, *6*(4), 353–369. https://doi.org/10.1016/j.jarmac.2017.07.008

Lewandowsky, S., & van der Linden, S. (2021). Countering Misinformation and Fake News Through Inoculation and Prebunking. *European Review of*

Social Psychology, *32*(2), 348–384.

https://doi.org/10.1080/10463283.2021.1876983

Lyons, T. (2018, May 21). *Hard Questions: What's Facebook's Strategy for Stopping False News? | Meta*.

https://about.fb.com/news/2018/05/hard-questions-false-news/

Maertens, R., Roozenbeek, J., Basol, M., & van der Linden, S. (2021). Long-term effectiveness of inoculation against misinformation: Three longitudinal experiments. *Journal of Experimental Psychology. Applied*, *27*(1), 1–16. https://doi.org/10.1037/xap0000315

Martel, C., Pennycook, G., & Rand, D. G. (2020). Reliance on emotion promotes belief in fake news. *Cognitive Research: Principles and Implications*, *5*(1), 47. https://doi.org/10.1186/s41235-020-00252-3

Martineau, P. (2019, May 2). Facebook Bans Alex Jones, Other Extremists—But Not as Planned. *Wired*.

https://www.wired.com/story/facebook-bans-alex-jones-extremists/

*Meedan launches collaborative effort to address misinformation on WhatsApp during Brazil's presidential election*. (2022, October 5).

https://meedan.com/post/meedan-launches-collaborative-effort-to-address-misinformation-on-whatsapp-during-brazils-presidential-election

Meredith, M., & Morse, M. (2015). *The Politics of the Restoration of Ex-Felon Voting Rights: The Case of Iowa* (SSRN Scholarly Paper No. 3872752). https://papers.ssrn.com/abstract=3872752

Modirrousta-Galian, A., & Higham, P. A. (2022). *How effective are gamified fake news interventions? Reanalyzing existing research with signal detection theory*.

Munger, K. (2017). Tweetment Effects on the Tweeted: Experimentally Reducing Racist Harassment. *Political Behavior*, *39*(3), 629–649. https://doi.org/10.1007/s11109-016-9373-5

Murrock, E., Amulya, J., Druckman, M., & Liubyva, T. (2018). Winning the war on

>state-sponsored propaganda: Results from an impact study of a Ukrainian

>news media and information literacy program. *Journal of Media Literacy*

>*Education*, *10*(2), 53–85.

Nieminen, S., & Rapeli, L. (2019). Fighting Misperceptions and Doubting

>Journalists' Objectivity: A Review of Fact-checking Literature. *Political*

>*Studies Review*, *17*(3), 296–309.

>https://doi.org/10.1177/1478929918786852

Nizamani, U. (2020). *Monetisation of Fake News in the Cyber Domain: A*

>*Roadmap for Building Domestic and International Cyber Resilience* (SSRN

>Scholarly Paper No. 3727833). https://papers.ssrn.com/abstract=3727833

Nyhan, B., Porter, E., Reifler, J., & Wood, T. J. (2020). Taking Fact-Checks

>Literally But Not Seriously? The Effects of Journalistic Fact-Checking on

>Factual Beliefs and Candidate Favorability. *Political Behavior*, *42*(3),

>939–960. https://doi.org/10.1007/s11109-019-09528-x

O'Callaghan, D., Greene, D., Conway, M., Carthy, J., & Cunningham, P. (2015).

>Down the (White) Rabbit Hole: The Extreme Right and Online

>Recommender Systems. *Social Science Computer Review*, *33*(4),

>459–478. https://doi.org/10.1177/0894439314555329

*Online Advertising Programme*. (2022, March 9). GOV.UK.

>https://www.gov.uk/government/consultations/online-advertising-programm

>e-consultation

*Online Civil Courage Initiative (OCCI)*. (n.d.). ISD. Retrieved 21 December 2022,

>from

>https://www.isdglobal.org/isd-programmes/online-civil-courage-initiative-oc

>ci/

*Online Safety Bill—Parliamentary Bills—UK Parliament*. (2022, December 20).

>https://bills.parliament.uk/bills/3137

Panayotov, P., Shukla, U., Sencar, H. T., Nabeel, M., & Nakov, P. (2022). *GREENER: Graph Neural Networks for News Media Profiling* (arXiv:2211.05533). arXiv. https://doi.org/10.48550/arXiv.2211.05533

Pantazi, M., Hale, S., & Klein, O. (2021). Social and Cognitive Aspects of the Vulnerability to Political Misinformation. *Political Psychology*, *42*(S1), 267–304. https://doi.org/10.1111/pops.12797

Pappas, S. (2022). Fighting fake news in the classroom. *Https://Www.Apa.Org*. https://www.apa.org/monitor/2022/01/career-fake-news

Paris, B., & Donovan, J. (2019). *Deepfakes and Cheap Fakes*. Data & Society. https://datasociety.net/library/deepfakes-and-cheap-fakes/

Pause before sharing, to help stop viral spread of COVID-19 misinformation. (2020, June 30). *UN News*. https://news.un.org/en/story/2020/06/1067422

Paynter, J., Luskin-Saxby, S., Keen, D., Fordyce, K., Frost, G., Imms, C., Miller, S., Trembath, D., Tucker, M., & Ecker, U. (2019). Evaluation of a template for countering misinformation—Real-world Autism treatment myth debunking. *PLOS ONE*, *14*(1), e0210746. https://doi.org/10.1371/journal.pone.0210746

Pennycook, G., Bear, A., Collins, E. T., & Rand, D. G. (2020). The implied truth effect: Attaching warnings to a subset of fake news headlines increases perceived accuracy of headlines without warnings. *Management Science*, *66*(11), 4944–4957.

Pennycook, G., Cannon, T. D., & Rand, D. G. (2018). Prior exposure increases perceived accuracy of fake news. *Journal of Experimental Psychology. General*, *147*(12), 1865–1880. https://doi.org/10.1037/xge0000465

Pennycook, G., Epstein, Z., Mosleh, M., Arechar, A. A., Eckles, D., & Rand, D. G. (2021). Shifting attention to accuracy can reduce misinformation online. *Nature*, *592*(7855), Article 7855. https://doi.org/10.1038/s41586-021-03344-2

Pennycook, G., McPhetres, J., Zhang, Y., Lu, J. G., & Rand, D. G. (2020). Fighting
    COVID-19 Misinformation on Social Media: Experimental Evidence for a
    Scalable Accuracy-Nudge Intervention. *Psychological Science*, *31*(7),
    770–780. https://doi.org/10.1177/0956797620939054

Pennycook, G., & Rand, D. G. (2022). Accuracy prompts are a replicable and
    generalizable approach for reducing the spread of misinformation. *Nature
    Communications*, *13*(1), Article 1.
    https://doi.org/10.1038/s41467-022-30073-5

Perez, S. (2019, July 2). Facebook News Feed changes downrank misleading
    health info and dangerous 'cures'. *TechCrunch*.
    https://techcrunch.com/2019/07/02/facebook-news-feed-changes-downran
    k-misleading-health-info-and-dangerous-cures/

Perez, S., & Hatmaker, T. (2020, November 5). Facebook blocks hashtags for
    #sharpiegate, #stopthesteal election conspiracies. *TechCrunch*.
    https://techcrunch.com/2020/11/05/facebook-blocks-sharpiegate-hashtag-e
    lection-conspiracies/

Portnoy, E. (2019, November 1). Why Adding Client-Side Scanning Breaks
    End-To-End Encryption. *Electronic Frontier Foundation*.
    https://www.eff.org/deeplinks/2019/11/why-adding-client-side-scanning-bre
    aks-end-end-encryption

Rauchfleisch, A., & Kaiser, J. (2021). *Deplatforming the Far-right: An Analysis of
    YouTube and BitChute* (SSRN Scholarly Paper No. 3867818).
    https://doi.org/10.2139/ssrn.3867818

Reis, J. C. S., Melo, P., Garimella, K., & Benevenuto, F. (2020). Can WhatsApp
    benefit from debunked fact-checked stories to reduce misinformation?
    *Harvard Kennedy School Misinformation Review*.
    https://doi.org/10.37016/mr-2020-035

Ribeiro, M. H., Jhaver, S., Zannettou, S., Blackburn, J., De Cristofaro, E.,

Stringhini, G., & West, R. (2021). Do Platform Migrations Compromise Content Moderation? Evidence from r/The_Donald and r/Incels. *Proceedings of the ACM on Human-Computer Interaction*, *5*(CSCW2), 1–24. https://doi.org/10.1145/3476057

Robertson, A. (2021, June 25). Google is warning users when its search results might be unreliable. *The Verge*. https://www.theverge.com/2021/6/25/22550430/google-search-results-changing-quickly-warning-breaking-news

Rogers, R. (2020). Deplatforming: Following extreme Internet celebrities to Telegram and alternative social media. *European Journal of Communication*, *35*(3), 213–229. https://doi.org/10.1177/0267323120922066

Roitero, K., Soprano, M., Portelli, B., De Luise, M., Spina, D., Mea, V. D., Serra, G., Mizzaro, S., & Demartini, G. (2021). Can the crowd judge truthfulness? A longitudinal study on recent misinformation about COVID-19. *Personal and Ubiquitous Computing*. https://doi.org/10.1007/s00779-021-01604-6

Roozenbeek, J., Freeman, A. L. J., & van der Linden, S. (2021). How Accurate Are Accuracy-Nudge Interventions? A Preregistered Direct Replication of Pennycook et al. (2020). *Psychological Science*, *32*(7), 1169–1178. https://doi.org/10.1177/09567976211024535

Roozenbeek, J., & Linden, S. van der. (2020). Breaking Harmony Square: A game that "inoculates" against political misinformation. *Harvard Kennedy School Misinformation Review*. https://doi.org/10.37016/mr-2020-47

Roozenbeek, J., Suiter, J., & Culloty, E. (2022a). *Countering Misinformation: Evidence, Knowledge Gaps, and Implications of Current Interventions* [Preprint]. PsyArXiv. https://doi.org/10.31234/osf.io/b52um

Roozenbeek, J., Suiter, J., & Culloty, E. (2022b). *Countering Misinformation: Evidence, Knowledge Gaps, and Implications of Current Interventions*.

PsyArXiv. https://doi.org/10.31234/osf.io/b52um

Roozenbeek, J., & van der Linden, S. (2019). Fake news game confers psychological resistance against online misinformation. *Palgrave Communications*, *5*(1), Article 1. https://doi.org/10.1057/s41599-019-0279-9

Roozenbeek, J., & van der Linden, S. (2022). How to Combat Health Misinformation: A Psychological Approach. *American Journal of Health Promotion*, *36*(3), 569–575. https://doi.org/10.1177/08901171211070958

Roozenbeek, J., van der Linden, S., Goldberg, B., Rathje, S., & Lewandowsky, S. (2022). Psychological inoculation improves resilience against misinformation on social media. *Science Advances*, *8*(34), eabo6254. https://doi.org/10.1126/sciadv.abo6254

Saltz, E., & Leibowicz, C. (2021). *Fact-Checks, Info Hubs, and Shadow-Bans: A Landscape Review of Misinformation Interventions*. Partnership on AI. https://partnershiponai.org/intervention-inventory/

Sardarizadeh, S., & Robinson, O. (2022, March 8). Ukraine invasion: False claims the war is a hoax go viral. *BBC News*. https://www.bbc.com/news/60589965

Shaar, S., Babulkov, N., Da San Martino, G., & Nakov, P. (2020). That is a Known Lie: Detecting Previously Fact-Checked Claims. *Proceedings of the 58th Annual Meeting of the Association for Computational Linguistics*, 3607–3618. https://doi.org/10.18653/v1/2020.acl-main.332

Shao, C., Ciampaglia, G. L., Varol, O., Yang, K.-C., Flammini, A., & Menczer, F. (2018). The spread of low-credibility content by social bots. *Nature Communications*, *9*(1), Article 1. https://doi.org/10.1038/s41467-018-06930-7

Shin, J., & Valente, T. (2020). Algorithms and Health Misinformation: A Case Study of Vaccine Books on Amazon. *Journal of Health Communication*, *25*(5),

394–401. https://doi.org/10.1080/10810730.2020.1776423

*Singapore Fake News Laws: Guide to POFMA (Protection from Online Falsehoods and Manipulation Act)*. (2022). SingaporeLegalAdvice.Com. https://singaporelegaladvice.com/law-articles/singapore-fake-news-protection-online-falsehoods-manipulation/

Smeros, P., Castillo, C., & Aberer, K. (2021). SciClops: Detecting and Contextualizing Scientific Claims for Assisting Manual Fact-Checking. *Proceedings of the 30th ACM International Conference on Information & Knowledge Management*, 1692–1702. https://doi.org/10.1145/3459637.3482475

Spangler, T. (2021, May 10). Facebook, Looking to Curb Misinformation, Is Starting to Prompt Users to Read Articles Before Sharing. *Variety*. https://variety.com/2021/digital/news/facebook-test-misinformation-read-article-1234969952/

Swire-Thompson, B., DeGutis, J., & Lazer, D. (2020). Searching for the Backfire Effect: Measurement and Design Considerations. *Journal of Applied Research in Memory and Cognition*, *9*(3), 286–299. https://doi.org/10.1016/j.jarmac.2020.06.006

Telford, T. (2019, February 22). Pinterest is blocking search results about vaccines to protect users from misinformation. *Washington Post*. https://www.washingtonpost.com/business/2019/02/21/pinterest-is-blocking-all-vaccine-related-searches-all-or-nothing-approach-policing-health-misinformation/

*The Debunking Handbook*. (2020, October 14). Skeptical Science. https://skepticalscience.com/the-debunking-handbook-redirect-page.shtml

*The Quarter Billion Dollar Question: How is Disinformation Gaming Ad Tech?* (2019). Global Disinformation Index. https://www.disinformationindex.org/

Traberg, C. S., Roozenbeek, J., & van der Linden, S. (2022). Psychological

inoculation against misinformation: Current evidence and future directions. *The ANNALS of the American Academy of Political and Social Science*, *700*(1), 136–151.

Twitter tests 'misleading' post report button for first time. (2021, August 18). *BBC News*. https://www.bbc.com/news/technology-58258377

Verstraete, M., Bambauer, D. E., & Bambauer, J. R. (2021). *Identifying and Countering Fake News* (SSRN Scholarly Paper No. 3007971). https://doi.org/10.2139/ssrn.3007971

Vorhaben, E. (2022). Technology Amplified Disinformation, Now It Must Demonetize It. *Chicago Policy Review*. https://chicagopolicyreview.org/2022/01/18/technology-amplified-disinformation-now-it-must-demonetize-it/

Vosoughi, S., Roy, D., & Aral, S. (2018). The spread of true and false news online. *Science*, *359*(6380), 1146–1151. https://doi.org/10.1126/science.aap9559

Vraga, E. K., & Bode, L. (2020). Defining Misinformation and Understanding its Bounded Nature: Using Expertise and Evidence for Describing Misinformation. *Political Communication*, *37*(1), 136–144. https://doi.org/10.1080/10584609.2020.1716500

Wardle, C., & Derakhshan, H. (2017). *Information disorder: Toward an interdisciplinary framework for research and policymaking*. Council of Europe.

West, D. M. (2017, December 18). How to combat fake news and disinformation. *Brookings*. https://www.brookings.edu/research/how-to-combat-fake-news-and-disinformation/

West, J. D., & Bergstrom, C. T. (2021). Misinformation in and about science. *Proceedings of the National Academy of Sciences*, *118*(15), e1912444117. https://doi.org/10.1073/pnas.1912444117

Yasseri, T., & Menczer, F. (2022). *Can crowdsourcing rescue the social marketplace of ideas?* (arXiv:2104.13754). arXiv. https://doi.org/10.48550/arXiv.2104.13754

Zerback, T., Töpfl, F., & Knöpfle, M. (2021). The disconcerting potential of online disinformation: Persuasive effects of astroturfing comments and three strategies for inoculation against them. *New Media & Society*, *23*(5), 1080–1098. https://doi.org/10.1177/1461444820908530

Zhou, X., Jain, A., Phoha, V. V., & Zafarani, R. (2020). Fake News Early Detection: A Theory-driven Model. *Digital Threats: Research and Practice*, *1*(2), 12:1-12:25. https://doi.org/10.1145/3377478